\documentstyle[preprint,prc,aps,floats,eqsecnum,epsf]{revtex}
\newcommand{\edc}{\end{document}}
\newcommand{\bb} {\begin{thebibliography}}
\newcommand{\eb}{\end{thebibliography}}
\newcommand{\bi}[1]{\bibitem{#1}}
\newcommand{\bc}{\begin{center}}
\newcommand{\ec}{\end{center}}
\newcommand{\disc}{\displaystyle}
\newcommand{\be}{\begin{equation}}
\newcommand{\ee}{\end{equation}}
\newcommand{\ba}{\begin{array}{l}   }
\newcommand{\lab}[1]{\label{#1}}
\newcommand{\ea}{\end{array}}
\newcommand{\Tr}{\mbox{Tr}}
\newcommand{\dsfrac}{\displaystyle\frac}
\newcommand{\ds} {\displaystyle}
\newcommand{\summa}{\ds\sum}
\newcommand{\re}[1]{(\ref{#1})}
\newcommand{\bra}[1]{\langle{#1}\vert}
\newcommand{\ket}[1]{\vert{#1}\rangle}
\newcommand{\matel}[3]{ \bra{#1}{#2}\ket{#3}}
\newcommand {\braket}[2]{ \bra{#1}{#2}\rangle }
\newcommand{\lbr}{\lbrack}
\newcommand{\rbr}{\rbrack}

\newcommand{\cl}{ {L}}
\newcommand{\cm}{\cal {M} }
\newcommand{\cmfi}   {{\cm}_{\mbox{fi}}  }
\newcommand{\cmfibar}   { {\overline{\cm}}_{\mbox{fi}}   }

\newcommand{\cmfione}{\cmfi^{(1)}}
\newcommand{\cmfitwo}{\cmfi^{(2)}}
\newcommand{\ci}{\cite}
\newcommand{\ch}[1]{\cosh\theta_{#1}}
\newcommand{\sh}[1]{\sinh\theta_{#1}}
\newcommand{\chsq}[1]{\cosh^2\theta_{#1}}
\newcommand{\shsq}[1]{\sinh^2\theta_{#1}}
\newcommand{\cs}[1]{\cos\theta_{#1}}
\newcommand{\sn}[1]{\sin\theta_{#1}}
\newcommand{\cssq}[1]{\cos^2\theta_{#1}}
\newcommand{\snsq}[1]{\sin^2\theta_{#1}}
\newcommand{\ta}{\tilde{a}}
\newcommand{\takrest}{\tilde{a}^\dagger}
\newcommand{\akrest}{{a}^\dagger}
\newcommand{\tc}{\tilde{c}}
\newcommand{\tckrest}{\tilde{c}^\dagger}
\newcommand{\ckrest}{{c}^\dagger}
\newcommand{\td}{\tilde{d}}
\newcommand{\tdkrest}{\tilde{d}^\dagger}
\newcommand{\dkrest}{{d}^\dagger}
\newcommand{\gamgcr}{\Gamma_{GCR}}
\newcommand{\lint} { L_{\mbox{int} } }

\newcommand{\betaq}{(\beta)}
\newcommand{\obeta}{0\betaq}
\newcommand{\tvacr}{\ket{\obeta}}

\newcommand{\bkrest}{{b}^\dagger}
\newcommand{\kone}{k_1}
\newcommand{\ktwo}{k_2}
\newcommand{\kuch}{k_3}
\newcommand{\km}{k_m}
\newcommand{\ckone}{ \ds{c_{\kone}   } }

\newcommand{\dktwo}{ \ds{d_{\ktwo}} }
\newcommand{\expon}[1]{{e}^{\disc{#1}}}
\newcommand{\mod}[1]{\vert{#1}\vert}
\newcommand{\pr}[1]{{#1}^\prime}
\newcommand{\pkone}{\pr{\kone}}
\newcommand{\pktwo}{ \pr{\ktwo} }
\newcommand{\pom}{\pr\omega}
\newcommand{\prn}{\pr{n} }
\newcommand{\prp}{\pr{p}  }
\newcommand{\prs}{\pr{s}}
\newcommand{\vecprp}{ \vec{\prp}      }
\newcommand{ \sqvec }[1] {  {  (\vec #1) }^2 }
\newcommand{\nwl}{\\[1mm]}
\newcommand{\sqrtd}{\sqrt{D_0}                  }
\newcommand{\lnmu}{\ln \frac{m^2}{\mu^2} }

\newcommand{\doublespace}
{\renewcommand{\baselinestretch}{1.6}\large\normalsize}
\doublespace
\begin{document}
\draft
\title{\Large\bf {Finite temperature amplitudes and reaction rates in
Thermofield dynamics.}\\
}
\author{
 A. Rakhimov \thanks
{Permanant address: Institute of Nuclear Physics, Tashkent,
    Uzbekistan (CIS)}
    and
 F.C. Khanna \thanks{E-mail: khanna@phys.ualberta.ca}
 }
\address{
Physics Department,     University of  Alberta
Edmonton,     Canada  T6G2J1 \\
 and\\
TRIUMF,     4004 Wesbrook Mall, Vancouver,     British Columbia,
Canada,     V6T2A3\\
}
\maketitle
\begin{abstract}
\medskip
\medskip
\medskip

We propose a method for calculating the reaction rates and transition
amplitudes of generic processes taking place in a many body system
 in equilibrium at temperature T.
The relationship of the scattering  and decay
amplitudes as calculated in Thermo Field Dynamics
 to the conventional techniques is established.
It is shown that in many cases the calculations are relatively easy
in TFD.
\end{abstract}
\medskip
\medskip
\newpage
\section{Introduction}
Recent experimental developments in heavy ion collision experiments
have lead to the formation of hadronic matter at finite temperature
with high densities. At collisions with relativistic heavy ions at
facilities  like  RHIC at  Brookhaven and beams at SPS, and eventually at
LHC, lead to the deconfinement of quarks from the confined state in hadrons
forming a quark gluon plasma. The phase transition is anticipated at
temperature of $T_c\approx 150$MeV - $250$ MeV. The quark gluon plasma
above the critical temperature $(T_c)$ and the hot and dense hadronic
matter below $T_c$ possess special problems to investigate their
properties at finite temperature and high densities.
We have to consider interaction of hadrons, the decay of mesons and
neutrons and finally the behavior of quark - quark interactions and
the screening effect in the quark gluon plasma (QGP).
Such a study requires the development of a formalism of quantum field
theory to deal with problems at finite temperature and high
densities.

It should be stressed that, developments to deal with hot and dense
nuclear matter and hot QGP would provide valuable tools to study
early universe as well as the interior of stars. Of course there is a
range of problems that would require time dependent treatment since
the system may be in a non - equilibrium state that is changing
rapidly with time. In general, therefore, it appears that both early
Universe and relativistic heavy ion reactions would require a
formalism that depends on time and temperature simultaneously.

 We wish to explore methods to calculate the reaction rates and decay
 rates in a medium at temperature $T$ using the real time finite
 temperature  field
 theory  - Thermo Field Dynamics (TFD) \ci{ashok}.
  This theory was  developed so that
 the time degree of freedom is not lost \ci{umetak}.
  It was realized that, an
 extention
 of  the usual field theory at zero temperature to finite
 temperatures would require doubling of the Hilbert space. At the same
 time  the operators have to be doubled such that the new set of
 operators, designated tilde operators, act on the second Hilbert space,
 tilde space. Later on it was interpreted that, the second Hilbert space
 acts like a heat bath that ensures the dynamical system to stay at
 constant temperature $T$. It is established that with TFD the Wick's
 theorem can be applied and other features like Ward - Takahshsi
 relations and Nambu - Goldstone theorem on spontaneously breaking
 symmetry can be established. Furthermore as in the case of the usual
  field
 theory at zero temperature,  Feynman rules for scattering amplitude
 and decay rates can be established. Therefore the combination of all
 these feature in TFD allows us to carry out calculations as in the
 case of $T= 0$ field theory. Some of the essential aspects
 such as  a definition of the vacuum and a definition of the creation and
 annihilation
 operators at finite temperature are given in the Appendix.
 It should be mentioned that for  systems in equilibrium
 there are two other methods available. The first is the imaginary
 time method due to Matsubara \ci{matsubara} that has been extended to
 quantum field theory by Jackiw and Dolan \ci{jackiwdolan}.
  The second method is the
 closed time path method due to Schwinger and Keldysh \ci{keldish}.
  Both these
 methods are essentually based on Green's function approach. Recently
 Niegawa \ci{niegawa} has developed a technique to use the closed time
 method to calculate the scattering amplitude and decay rates. We'll
 compare this method with that of TFD later on.

 It should be stressed that what has been done is to write down
 the decay amplitudes by using the method of Cutkosky to find the
 imaginary part of an amplitude. These rules are very helpful in our
 investigation.

Recently Jeon and Ellis [6] have developed a multiple scattering
expansion of the self energy at finite temperature in the imaginary
time method. They have calculated the response function in this
approach. Brandt et al [6] have calculated the retarded thermal
Green's function and forward scattering amplitude at two loops. Using
time-dependent two- and three-point functions in scalar theory they
have calculated i) the forward scattering amplitude of two on-shell
thermal particles and ii) forward scattering amplitude of a single
on-shell thermal particle. The calculations include one-loop self
energy and vertex corrections that are retarded but are at zero
temperature. The present work attempts to calculate scattering and
decay amplitudes at finite temperatures.

The paper is organised as follows.
In sect. II we briefly present basic TFD formulas, which
will be nessesary in further calculations. In sect. III we
describe  our
basic formalism. In sections  IV, V and VI we consider
 decay and scattering
procecces respectively. In sect VII we consider loop contributions.
The summary and conclusions are given
 in sect VIII.
\section{Thermal fields and S - matrix in TFD.}
The Thermo Field Dynamics is a real time operator formalism
of quantum field theory at finite temperature. Any physical state can
be constructed from a temperture dependant vacuum
$\ket{\obeta}$ which is a pure state.
The main feature of TFD is that,
 the thermal average of \underline{any operator}
 $A$ is equal to its temperature dependent vacuum expectation
 value with the:
vacuum, $\ket{0(\beta)}$, being obtained from the usual vacuum by
a  Bogoliubov transformation. Therefore, we have
\be
<A>=\matel{0(\beta)}{ A}{0(\beta)}
\lab{ampl2.1}
\ee
where $\beta \equiv 1/k_{B}T$ with $k_{B}$ being the Boltzmann
constant.
Doubling of degrees of freedom of all fields is required. This is
achieved
through a "tilde" operation: to each zero temperature field $
\phi(x)$ a doublet of fields $(\phi(x),\tilde\phi(x))$ is attached, the
dynamics of which is controlled by a thermal Lagrangian
\be
\hat {\cl}=\cl (\phi)-\tilde {\cl}  (\phi)=\cl (\phi)-{\cl}^{*}
(\tilde{\phi})
\lab{ampl2.2}
\ee
The temperature dependent vacuum state  $\ket{0(\beta)}$ is
annihilated by the  temperature dependent physical annihilation operators
\be
\ba
a_{k}(\beta)\ket{0(\beta)}=c_k(\beta)\ket{0(\beta)}=d_k(\beta)
\ket{0(\beta)}=0
\nwl
\tilde a_k(\beta) \ket{0(\beta)}=\tc_k(\beta)\ket{0(\beta)}=
\td_k(\beta) \ket{0(\beta)}=0
\lab{ampl2.3}
\ea
\ee
which are obtained through Bogoluibov transformation from usual
annihilation, $a_k$,  and creation operators,
$\akrest_k$. For a Bose system
\be
\ba
a_k=\ch{k}a_k(\beta)+\sh{k}\takrest_k\betaq, \quad
\akrest_k=\ch{k}\akrest_k(\beta)+\sh{k}\ta_k\betaq,
\nwl
\takrest_k=\sh{k}a_k(\beta)+\ch{k}\takrest_k\betaq,\quad
\ta_k=\sh{k}\akrest_k(\beta)+\ch{k}\ta_k\betaq,
\lab{ampl2.4}
\ea
\ee
where
\be
\ba
\shsq{k}=n_B(k)=1/(\expon{\beta\mod{k_0}}-1), \quad
 \ch{k}=\expon{\beta
k_0/2}\sh{k} .
\lab{ampl2.5}
\ea
\ee
with $k_0$ being the energy associated with the four - vector, $k$.
The creation and annihilation operators satisfy ordinary commutation
relations
\be
\ba
\lbr a_k,\akrest_p\rbr _-=\lbr \ta_k,\takrest_p\rbr _-=\lbr
a_k\betaq,\akrest_p\betaq\rbr _-=
\lbr \ta_k\betaq,\takrest_p\betaq\rbr _-=\delta(\vec{k}-\vec{p})
\lab{ampl2.6}
\ea
\ee
and all other commutators vanish.

For fermions, with the creation, $\ckrest_{p,s}    $ and
 $ \dkrest_{p,s} $ , and annihilation, $ c_{p,s}  $ and $ d_{p,s} $
operators the Bogoliubov transformations lead to the relations
\be
\ba
c_{p,s}=\cs{+p}c_{p,s}\betaq+i\sn{+p}\tckrest_{p,s}\betaq ,\quad
d_{p,s}=\cs{-p}d_{p,s}\betaq+i\sn{-p}\tdkrest_{p,s}\betaq ,\quad
\nwl
\tckrest_{p,s}= i\sn{+p}c_{p,s}\betaq + \cs{+p}\tckrest_{p,s}\betaq ,\quad
\tdkrest_{p,s}= i\sn{-p}d_{p,s}\betaq +
\cs{-p}\tdkrest_{p,s}\betaq,
\nwl
\ckrest_{p,s}=\cs{+p}\ckrest_{p,s}\betaq-i\sn{+p}\tc_{p,s}\betaq ,\quad
\dkrest_{p,s}=\cs{-p}\dkrest_{p,s}\betaq-i\sn{-p}\td_{p,s}\betaq ,\quad
\nwl
\tc_{p,s}= -i\sn{+p}\ckrest_{p,s}\betaq + \cs{+p}\tc_{p,s}\betaq, \quad
\td_{p,s}=- i\sn{-p}\dkrest_{p,s}\betaq + \cs{-p}\td_{p,s}\betaq,
\lab{ampl2.7}
\ea
\ee
where
\be
\ba
\snsq{\pm p }=1/(1+\expon{\beta\mod{p_0\mp\mu}}), \quad
\cs{\pm p}=\expon{\beta(p_0\pm\mu)/2}\sn{\pm p}
\nwl
\snsq{+p}=n_F(p), \quad
\snsq{-p}=\bar{n}_F(p) .
\lab{ampl2.8}
\ea
\ee
Here $p_0$ is the energy associated with the four - vector, $p$
and $\mu$ is the chemical potential.
The anticommutation relations for creation and annihilation operators (c-
for particles, and d - for antiparticles) are similar to those  at zero
temperature:
\be
\ba
\lbr c_{\prp \prs}   \ckrest_{p,s}\rbr _{+}=\lbr d_{\prp
\prs}\dkrest_{p,s}\rbr _{+}=
\lbr c_{\prp \prs} \betaq  \ckrest_{p,s}\betaq\rbr _{+}=\lbr d_{\prp \prs}
\betaq
\dkrest_{p,s}\betaq\rbr _{+}=
\delta(\vec{p}-\vecprp)\delta_{s\prs}
\nwl
{\lbr \tc_{\prp \prs}   \tckrest_{p,s}\rbr }_{+} =
{\lbr \td_{\prp \prs}
\tdkrest_{p,s}\rbr }_ {+} =
{\lbr \tc_{\prp \prs} \betaq  \tckrest_{p,s}\betaq\rbr }_{+} =
{\lbr \td_{\prp \prs} \betaq
\tdkrest_{p,s}\betaq\rbr }_ {+} =\delta(\vec{p}-\vecprp)\delta_{s\prs}
\lab{ampl2.9}
\ea
\ee
All other anti - commutators are zero.
In order to define the reaction rates and decay widths,
we follow the prescription due to Feynman
\ci{feynman}, which is best stated by
Dyson \ci{dyson}, and this involves writing down an operator:
\be
\ba
H_F(x_0)=
\summa_{n=0}^{\infty}
\dsfrac{(-i)^n   }{n!}\ds\int_{-\infty}^{+\infty} dx_1 dx_2\dots dx_n
T\lbr H^e(x_0)H_I(x_1)H_I(x_2)\dots H_I(x_n)\rbr ,
\lab{ampl2.10}
\ea
\ee
Then choosing $H^e(x_0)=1$  this reduces to an expression that is
evaluated between an initial and a final state. Therefore we get
an expression that looks like and is equal to an $S$ matrix
\be
\ba
S=\summa_{n=0}^{\infty}S^n=
\summa_{n=0}^{\infty}
\dsfrac{(-i)^n   }{n!}\ds\int dx_1 dx_2\dots dx_n
T\lbr H_I(x_1)H_I(x_2)\dots H_I(x_n)\rbr ,
\lab{ampl2.11}
\ea
\ee
For the total transition in TFD we have
\be
\ba
\hat S=\summa_{n=0}^{\infty}\hat S^n=\summa_{n=0}^{\infty}
\dsfrac{(-i)^n}{n!}\ds\int dx_1 dx_2\dots
 dx_n T\lbr \hat H_I(x_1)\hat H_I(x_2)\dots \hat H_I(x_n)\rbr ,
\lab{ampl2.12}
\ea
\ee
where $\hat H_I(x)=H_I(x)-\tilde H_I(x)$  \ci{saito}.
Particularly,  at  the tree level
\be
\hat S\approx 1-i\ds\int d^4x (H_I(x)-\tilde H_I(x))
\lab{ampl2.13}
\ee

Now in these expressions $H_I(x)$ and $\tilde H_I(x)$
may be introduced explicitly and then Wick's theorem is used as in
the case of $T=0$ field theory to obtain all the contributions in
perturbation theory. The evaluation proceeds by using a set of Feynman
rules for evaluation of a reaction rate and decay widths.
It is clear that for operators appearing in $H_I(x)$ and
$\tilde H_I(x)$, the transformation of boson and Fermion operators
given by Eq.s \re{ampl2.4} and \re{ampl2.7} respectively has to be
used to bring in the temperature dependent factors. The one particle
Green's functions are given explicitly in the Appendix.
\section{ Calculation of reaction rates}
Let us consider the  process
\be
p_1+p_2+\dots +p_n\rightarrow \prp_1+\prp_2+\dots +\prp_m
\lab{ampl3.1}
\ee
At zero   temperatute, $T=0$,  the amplitude of this process
 can be calculated by usual Feynman rules:
\be
\matel{f}{\hat S}{i}=\summa_{n=0}^{\infty} \matel{f}{\hat S^{(n)} }{i}
\lab{ampl3.2}
\ee
where $\ket {i} =\akrest_{p_1}\akrest_{p_2}\dots \akrest_{p_n}\ket{0}$,
$\ket {f} =\akrest_{\prp_1}\akrest_{\prp_2}\dots
\akrest_{\prp_m}\ket{0}$ and $\ket{0}$ is an ordinary  vacuum
state: $a_p\ket{0}=0$.
    In the present article we propose that the amplitude and,
       hence the cross section, of process \re{ampl3.1}
 can be directly calculated perturbatively using
 Eq.s \re{ampl2.12}   at $T\not= 0$ also.
Namely, we propose that

1) The transition amplitudes  are  determined from Eq.
 \re{ampl2.12} where for the process \re{ampl3.1}
  \,\,\,  we have $\ket {i}=\akrest_{p_1}\betaq\akrest_{p_2}\betaq
  \dots \akrest_{p_n}\betaq\ket{0(\beta)}$ and
$\ket {f}=\akrest_{\prp_1}\betaq
\akrest_{\prp_2}\betaq\dots
 \akrest_{\prp_m}\betaq \ket{0(\beta)}      $ .

2) The phase space factors, relating an amplitude
 to the rate are the same as in zero temperature case.
 For example, the differential cross section for the process
 $p_1+p_2\rightarrow\prp_1+\prp_2+\dots +\prp_m$
is  \ci{mandlshaw}
\be
d\sigma=(2\pi)^4\delta^4(\prp_1+\prp_2+\dots +\prp_m -p_1-p_2      )
\dsfrac{1}{ 4E_1E_2v_{\mbox{rel}} } \prod_l(2m_l)
\prod_{i=1}^{m}\dsfrac{ d\vec {\prp}_i }{ (2\pi)^32E_{i}^{\prime}     }
\mod{ \cmfi }^2
\lab{ampl3.3}
\ee
where $E_p=\sqrt{m_{i}^{2}+ \sqvec{p}   }$ and $ v_{\mbox{rel}}$ is
the relative velocity of the two initial particles with momenta
$p_1$ and $p_2$.
 The amplitude is related to $ \hat S$ matrix
in manner
similar  to the procedure carried out at zero temperature
\be
\matel{f}{\hat S}{i}=i(2\pi)^4\delta^4(P_f-P_i)\cmfi
\prod_{ext}\lbr\dsfrac{m}{VE}\rbr^{1/2}
\prod_{ext}\lbr\dsfrac{m}{V\omega}\rbr^{1/2}.
\lab{ampl3.4}
\ee
Here $P_i$ and $P_f$ are the total four momenta in the initial and
 final states, and the products extend over all external
 fermions and bosons, $E$ and $\omega$ being the energies
  of the individual external fermions and bosons respectively,
  and $V$ is the volume. We propose that,
in TFD this definition  still holds.
\section{Decay of particles}
 We know that at $T=0$ the decay
rates $\Gamma$ can be calculated either by studying
 discontinuities  of the self energy
of a decaying particle (Cutkosky rules \ci{cutkosky})
 or by direct calculation of
the amplitude perturbatively (Feynman rules). Obviously these two
 methods must lead to the same  result.

For finite temperature the Cutkosky rules were generalized by
Kobes and Semenoff \ci{kobes}. In the next section we shall
 calculate the  decay of a particle   at finite temperature \\
a) by using generalized Cutkosky rules (GCR); and also \\
b) directly as  a square of module of elements of $\hat S$ matrix,
 i.e. by using the  above method.\\
It will be established  that, these two
 methods give the same result for $\Gamma$.
We shall present these to establish their equivalence.
\nwl
\nwl
{\bf A. } $\sigma(k)\rightarrow \pi^0(\kone)+\pi^0(\ktwo)$
\nwl

Firstly, we calculate the decay rate of sigma meson
  into two pions  (we'll omit isospin indices for simplicity) in real time
formailsm,
  using GCR. The interaction Lagrangian
  is $\lint=\dsfrac{\lambda}{2}\sigma(x)\pi^2(x)$.
The decay rate of a boson with mass $M$ and four
 momentum $k=(\omega,\vec k)$ with
 $ \omega=\sqrt{ M^{2}+ \sqvec{k} }$, is related to the  self energy
\ci{fujimoto} $\bar\Sigma(\omega)$ by
\be
\gamgcr(w)=-\dsfrac{1}{\omega}Im \bar\Sigma(\omega)=
-\dsfrac{(\expon{\beta\omega}-1)}{\omega(\expon{\beta\omega}+1)}
Im\Sigma_{11}(k)
\lab{ampl4.1}
\ee
where
\be
i\Sigma_{11}(k)=\lambda^2\ds\int \dsfrac{dp^4 }{(2\pi)^4}
i\Delta_{11}^{0}(p)i\Delta_{11}^{0}(p-k)
\lab {4.2}
\ee
The GCR \ci{kobes}, as   illustrated in Fig. 1  give:
\be
2Im\Sigma_{11}(k)=\lambda^2\ds\int \dsfrac{dp^4 }{(2\pi)^4}
\lbr i\Delta^{+}(p)i\Delta^{-}(p-k)+i\Delta^{-}(p)i\Delta^{+}(p-k)\rbr
\lab {4.3}
\ee
where $i\Delta^{\pm}(p)=2\pi\lbr \Theta(\pm p_0)+n_B(p)\rbr \delta(p^2-m^2)$.
The two terms in the integrand are related by
\be
i\Delta^{\pm}(p)=i\expon{\pm\beta p_0}\Delta^{\mp}(p)
\lab{ampl4.4}
\ee
from \re{ampl4.1} -\re{ampl4.4} we get:
\be
\ba
\gamgcr(w)=\dsfrac{(\expon{\beta\omega}-1)
(\expon{-\beta\omega}+1)\lambda^2}          {2\omega(\expon{\beta\omega}+1)}
\ds\int \dsfrac{dp^4 }{(2\pi)^4}(2\pi)^2 \delta(p^2-m^2)\delta(
(p-k)^2-m^2)\times
\nwl
\times\lbr\Theta( p_0)+n_B(p_0)\rbr \lbr\Theta(
-p_0+\omega)+n_B(p_0-\omega)\rbr
\lab{ampl4.5}
\ea
\ee

We shall proceed and evaluate $\gamgcr$ explicitly.
 Note, however that, at finite temperature Lorentz invariance is
 lost,  and hence, the decay rate will no longer be Lorentz
 invariant,
   but dependent on the reference frame. In the following
   examples we will choose the rest frame of the decaying particle,
as the reference frame,     that is we set $\vec k=0$.
First, we consider the product of the mass shell $\delta$ functions. With
$k=(M,0,0,0)$
the compatible zeros are
easily found to be $p_0=\omega_p=M/2.$ Hence the delta functions reduce to :
\be
\delta (p^2-m^2)\delta( (p-k)^2-m^2)=\dsfrac{1}{4M\omega_p}
\delta(p_0-\omega_p)\delta( \omega_p-M/2)
\lab{ampl4.6}
\ee
and fix the momentum dependence of the integrand completely.

The thermal factors in \re{ampl4.5}  can be reduced on mass shell:
$p_0=M/2$, $\omega=M$
\be
\ba
\dsfrac{(\expon{\beta\omega}-1)
(\expon{-\beta\omega}+1)}{(\expon{\beta\omega}+1)}
\lbr\Theta( p_0)+n_B(p_0)\rbr \lbr\Theta( -p_0+\omega)+n_B(p_0-\omega)\rbr =
\nwl
\quad
=(1-\expon{-\beta\omega})\lbr1+n_B(p_0)\rbr \lbr1+n_B(p_0-\omega)\rbr =
(1-\expon{-\beta M})(1+n_B(M/2))^2
\lab{ampl4.7}
\ea
\ee

now inserting \re{ampl4.6}, \re{ampl4.7} into \re{ampl4.5} we get:
\be
\ba
\dsfrac{ \gamgcr(T\neq0) }{ \Gamma(T=0) }=\dsfrac{ (1+n_B(M/2))^2 }{
1+n_B(M) }=
(1+n_B(M/2))^2-n_{B}^{2}(M/2)
\lab{ampl4.8}
\ea
\ee
Here we used the simple relations:
\be
\ba
\expon{-\beta M}=\expon{-\beta M/2}\expon{-\beta M/2}
\nwl
\expon{-\beta\omega}=n_B(\omega)/(1+n_B(\omega))
\lab{ampl4.9}
\ea
\ee
which will be useful for further calculations.

Now we calculate the rate $\Gamma(\sigma\rightarrow \pi\pi)$
directly.
Because of the doubling of degrees of freedom the whole interaction
Lagrangian
will be
$\hat\lint=\lambda\sigma(x)\pi^2(x)-\lambda\tilde\sigma(x){\tilde\pi}^2(x)$.
The initial and final states are $\ket{i}=\akrest_k\betaq\ket{0\betaq}$ and
$\ket{f}=\bkrest_{\kone}\betaq \bkrest_{\ktwo}\betaq    \ket{0\betaq}$
where operators $a\betaq$ and
$b\betaq$ stand for $\sigma$ and $\pi$ fields respectively.
At  the tree level \re{ampl2.14} the matrix element of $\hat S$ matrix is
given by
\be
\matel{f}{\hat S}{i}=i\lambda \ds\int dx \matel{0\betaq}
{b_{\kone}\betaq b_{\ktwo}\betaq\lbr\sigma(x)\pi^2(x)-\tilde
\sigma(x){\tilde\pi}^2(x)\rbr \akrest_k\betaq}{0\betaq}
\lab{ampl4.10}
\ee
where the boson fields $\sigma(x)$ and $\pi(x)$
are defined  as in  eq. \re{amplA22}.
Using Bogoluibov transformations \re{ampl2.4}
 and commutation relations \re{ampl2.6}
it is easy to show that:
\be
\ba
\matel {0\betaq}{\sigma(x)\akrest_k\betaq}{0\betaq}=
\expon{-ikx}\ch{k},
\nwl
\matel {0\betaq}{\tilde\sigma(x)\akrest_k\betaq}{0\betaq}=
\expon{-ikx}\sh{k},
\nwl
\matel {0\betaq}{b_{\kone}\betaq b_{\ktwo}\betaq \pi^2(x) }{0\betaq}=
\expon{i(\kone+\ktwo)x}\ch{\kone}\ch{\ktwo}
\nwl
\matel {0\betaq}
 { b_{\kone} \betaq b_{\ktwo} \betaq {\tilde\pi}^2(x) }
 {0\betaq}=
\expon{i(\kone+\ktwo)x}
\sh{\kone}\sh{\ktwo}.
\lab{ampl4.11}
\ea
\ee
>From Eqs \re{ampl4.10}, \re{ampl4.11} and \re{ampl2.12} we obtain the
amplitude:
\be
\cmfi (T)=\lambda
\lbr\ch{k}\ch{\kone}\ch{\ktwo}-\sh{k}\sh{\kone}\sh{\ktwo}\rbr .
\lab{ampl4.12}
\ee
The decay rate,
 can be  calculated  directly from \re{ampl4.12}
  without introducing any additional thermal factor:
\be
\ba
\Gamma(\omega)=\dsfrac{1}{2\omega}
\ds\int \dsfrac{ d\vec \kone d\vec \ktwo (2\pi)^4\delta(k-\kone-\ktwo)
\mod{\cmfi(T)}^2 } {(2\omega_1) (2\omega_2)(2\pi)^3(2\pi)^3
}=
\nwl
\nwl
=\dsfrac{\lambda^2}{32\omega\pi^2}
\ds\int \dsfrac{ d\vec \kone d\vec \ktwo \delta^4(k-\kone-\ktwo)
W_B(\omega,\omega_1,\omega_2)} {\omega_1 \omega_2 }\equiv
\dsfrac{\lambda^2  I_B(T)}{32\omega\pi^2}
\lab{ampl4.13}
\ea
\ee
where
\be
W_B(\omega,\omega_1,\omega_2)=
\lbr\ch{k}\ch{\kone}\ch{\ktwo}-\sh{k}\sh{\kone}\sh{\ktwo}     \rbr ^2,
\lab{ampl4.14}
\ee
 $\omega_i=\sqrt{\sqvec{k_i}+m^2}$ and $\omega=\sqrt{\sqvec{k}+M^2}$.
Now using the properties of the hyperbolic functions and Eqs \re{ampl2.5},
and \re{ampl4.9}
  this may be further simplified as  follows:
\be
\ba
\delta(\omega-\omega_1-\omega_2)W_B(\omega,\omega_1,\omega_2)=
n_1n_2n_\omega\lbr\expon{\beta(\omega+\omega_1+\omega_2)/2}-1\rbr ^2
\delta(\omega-\omega_1-\omega_2)=
\nwl
=n_1n_2n_\omega(\expon{\beta\omega}-1)(\expon{\beta(\omega_1+\omega_2}-1)
\delta(\omega-\omega_1-\omega_2)=
\nwl
=n_1n_2n_\omega\lbrace\dsfrac{ 1+n_\omega}{n_\omega}-1\rbrace
\lbrace\dsfrac{(1+n_1)}{n_1}\dsfrac{(1+n_2)}{n_2}-1\rbrace
\delta(\omega-\omega_1-\omega_2)=
\nwl
=(1+n_1+n_2)\delta(\omega-\omega_1-\omega_2)
\lab{ampl4.15}
\ea
\ee
where for simplicity we denote
$n_i\equiv n_B(k_i)$, $ (i=1,2), $ and
$n_\omega\equiv n_B(k)$.

The integral $I_B(T)$ in Eq. \re{ampl4.13}
 can be explicitly calculated in the rest frame
  of the decaying particle: $\omega=M$,
   $\vec k=0$, $\omega_i=\sqrt{\sqvec{k_i}+m^2}
   \equiv\sqrt{\sqvec{q}+m^2}=\omega_q$:
\be
\ba
I_B(T)\equiv
\ds\int \dsfrac{ d\vec \kone d\vec \ktwo \delta^4(k-\kone-\ktwo)
W_B(\omega,\omega_1,\omega_2)} {\omega_1 \omega_2 }=
\nwl
\quad
\quad
\quad
=4\pi\ds\int \dsfrac{dq q^2
\delta(2\omega_q-M)W_B(M,\omega_q,\omega_q)}{\omega_{q}^{2}}=
8\pi\sqrt{ (1/4-(m/M)^2 }W_B( M,M/2,M/2)
\lab{ampl4.16}
\ea
\ee
using \re{ampl4.15} and  \re{ampl4.16} in \re{ampl4.13} we finally obtain:
\be
\ba
\dsfrac{\Gamma(T\neq0)}{\Gamma(T=0)}=W_B(M,M/2,M/2)=(1+2n_B(M/2))
\lab{ampl4.17}
\ea
\ee
this formulae is the same as obtained by the  GCR method eq. \re{ampl4.8}
\nwl
\nwl
{\bf B. } $H(k)\rightarrow \expon{-}(\kone)+\expon{+}(\ktwo)$
\nwl

As a second example we consider the decay of a boson  into two fermions,
namely, the decay of Higgs boson into electron - positron pair.
The decay rate of this process
 has been calculated in detail using GCR by Keil
\ci{keil}. The final tree level result is :
\be
\ba
\dsfrac{\gamgcr(T\neq0)}{\Gamma(T=0)}=
\dsfrac{(1-n_F(M/2))(1-\bar n_F(M/2))}{1+n_B(M)}
\lab{ampl4.18}
\ea
\ee
which is obtained in the rest frame of $H$ boson with a mass $M$.

Now we evaluate the  same rate by direct calculation
of the amplitude.
The interaction Lagrangian is

\be
\ba
 \hat{L}_{\mbox{int}}  =\lint-\tilde\lint ,
 \quad  \lint=-igH(x)\bar\psi(x)\psi(x), \quad
  \tilde\lint=+ig\tilde H(x)\bar{\tilde\psi}(x)\tilde\psi(x) .
\lab{ampl4.19}
\ea
\ee
The matrix element of $\hat S$ matrix from an initial state
 $\ket{i}=\akrest_k\betaq\ket{\obeta}$ to the final state
$\ket{f}=\ckrest_{\kone}\betaq \dkrest_{\ktwo}\betaq    \ket{\obeta}$
at  the tree level is
\be
\ba
\matel{f}{\hat S}{i}=i (-ig) \ds\int dx^4 \matel{ i  } {
\lbr H(x)\bar\psi(x)\psi(x)
+\tilde H(x)\bar{\tilde\psi}(x)\tilde\psi(x)          \rbr }{ f}
\lab{ampl4.20}
\ea
\ee
where we omit the spin indices for simplicity.
As in previous example, Eq. \re{ampl4.11}, we have
\be
\ba
\matel {f}{H(x)}{i}=\expon{-ikx}\ch{k} , \quad
\matel {f}{\tilde H(x)}{i}=\expon{-ikx}\sh{k} .
\ea
\lab{ampl4.21}
\ee
 For the fermions, using \re{amplA23} in  \re{ampl4.20}
  we have
\be
\ba
\matel{f}{ \bar\psi(x)\psi(x) }{i}=\ds\int d\vec p
d
\vecprp N_pN_\prp
 \bra{0\betaq}
 \ckone\betaq \dktwo\betaq
\lbr\ckrest_{p}\bar{u} (p)\expon{ipx}+d_{p}\bar{v}(p)\expon{-ipx}\rbr \times
\nwl
\times \lbr c_{\prp} u(\prp)\expon{-i\prp x}+\dkrest_{\prp}
v(\prp)\expon{i\prp
x}\rbr
\ket{ 0\betaq }
\lab{ampl4.22}
\ea
\ee
and
\be
\ba
\matel{f}{ \bar{\tilde\psi}(x)\tilde\psi(x)    }{i}=\ds\int d\vec p
d
\vecprp N_pN_\prp
\bra{0\betaq}
 \ckone\betaq \dktwo\betaq
\lbr\tckrest_{p}\bar{\tilde u}(p)\expon{-ipx}+\td_{p}\bar{\tilde
v}(p)\expon{ipx}\rbr \times
\nwl
\times
\lbr\tc_{\prp}\tilde u(\prp)\expon{i\prp x}+
\tdkrest_{\prp}\tilde v(\prp)\expon{-i\prp x}\rbr
\ket{0\betaq}
\lab{ampl4.23}
\ea
\ee
   The only "surviving" term in \re{ampl4.22}
  is that one which includes
\be
\ba
\bra {0\betaq} \ckone\betaq \dktwo\betaq
\ckrest_{p}\dkrest_{\prp} \ket{0\betaq}=\bra {0\betaq}
  \ckone\betaq \dktwo\betaq \lbr\cs{+p} \ckrest_{p}\betaq-i
  \sn{+p}\tc_p\betaq\rbr =
\nwl
\quad
=\lbr\cs{-\prp} \dkrest_{\prp}\betaq- i
\sn{-\prp}\td_{\prp}\betaq\rbr \ket{0\betaq} =
\matel {0\betaq} {\ckone\betaq \dktwo\betaq \ckrest_{p}\betaq
\dkrest_{\prp}\betaq}{0\betaq}\times
\nwl
\times\cs{+p}\cs{-\prp}\cs{+\kone}\cs{-\ktwo}\delta(\vec \kone-\vec p)
\delta(\vec \ktwo-\vecprp) .
\lab{ampl4.24}
\ea
\ee
Similarly, the nonzero contribution to \re{ampl4.23} involves
\be
\ba
\bra {0\betaq} \ckone\betaq \dktwo\betaq \td_{p}\tc_{\prp}\ket{0\betaq}
=\bra {0\betaq}  \ckone\betaq \dktwo\betaq
\lbr\cs{-p} \td_p\betaq- i \sn{-p}\dkrest_p\betaq\rbr \times
\nwl
\times
\lbr {\cs{+ {\prp} }} {\tc_{\prp}}\betaq-i\sn{+{ \prp} }\ckrest_{\prp}
\betaq \rbr \ket{0\betaq}=
\nwl
=-\bra {0\betaq} \ckone\betaq \dktwo
\betaq
\ckrest_{\prp}\betaq\dkrest_p\betaq\ket{0\betaq}\sn{-p}\sn{+\prp}
\nwl
=-\sn{+\kone}\sn{-\ktwo}\delta(\vec \kone-\vecprp)
\delta(\vec\ktwo-\vec p) .
\lab{ampl4.25}
\ea
\ee

The decay of the process is given by
a general formula of the zero temperature field theory:
\be
\ba
\Gamma(\omega)=\dsfrac{1}{2\omega}
\ds\int \dsfrac{ d\vec \kone d\vec \ktwo (2\pi)^4\delta^4(k-\kone-\ktwo)
(2m)^2\mod{\cmfi (T)}^2 } { (2\omega_1)
 (2\omega_2)(2\pi)^3(2\pi)^3 }
\lab{ampl4.26}
\ea
\ee
where the transition amplitude is  given by Eqs \re{ampl3.4},
and for the present case it has the form
\be
\cmfi(T)=(-ig)\lbr \cs{\kone}\cs{-\ktwo}\ch{\omega}\bar u(\kone)v(\ktwo)-
\nwl
\sn{\kone}\sn{-\ktwo}\sh{\omega}\bar{\tilde v}(\ktwo)\tilde
u(\kone)\rbr .
\lab{ampl4.27}
\ee
Then Eq. \re{ampl4.26} can be fiurther simplified by using
\be
\ds\sum_{\mbox{spins}}\mod{\cmfi (T)}^2    =
g^2\lbr
\cs{\kone}\cs{-\ktwo}\ch{\omega}-\sn{\kone}\sn{-\ktwo}\sh{\omega}\rbr ^2
Tr\dsfrac{ (\not {\kone} +m)}{2m}\dsfrac{(\not{\ktwo}+m)}{2m}
\lab{ampl4.28}
\ee
and
\be
\ba
\delta(\omega-\omega_1-\omega_2)
\lbr \cs{\kone}\cs{-\ktwo}\ch{\omega}-
\sn{\kone}\sn{-\ktwo}\sh{\omega}\rbr ^2=
\nwl
=\delta(\omega-\omega_1-\omega_2)
\cssq{\kone}\cssq{-\ktwo}\chsq{\omega}(1-\expon
{-\beta(\omega+\omega_1+\omega_2)/2})^2=
\nwl
=\delta(\omega-\omega_1-\omega_2)
(1-n_F(\kone)(1-\bar n_F(\ktwo)
(1-\expon{-\beta\omega})
\equiv\delta(\omega-\omega_1-\omega_2)W_F(\omega,\omega_1,\omega_2) .
\lab{ampl4.29}
\ea
\ee
Now, inserting \re{ampl4.27} -\re{ampl4.29} into \re{ampl4.26} and
performing integration in the rest
 frame of the decaying boson we  obtain
\be
\ba
\dsfrac{\Gamma(T\neq0)}{\Gamma(T=0)}=
W_F(M,M/2,M/2)=(1-n_F(M/2))(1-\bar n_F(M/2))
(1-\expon{-\beta M})
\lab{ampl4.30}
\ea
\ee
which is exactly the same as  Eq. \re{ampl4.18} given by GCR,  since $
(1-\expon{- \beta \omega})=(\chsq{\omega})^{-1}=(1+n_B(\omega))^{-1}$ .
\nwl
\nwl
{\bf C.}   $\Phi(k)\rightarrow \phi(\kone)+\phi(\ktwo)+\phi(\kuch)$
\nwl

The next example is the decay rate of a boson -$\Phi(x)$ into three bosons -
$\phi(x)$ with the interaction Lagrangian
$\lint=-\dsfrac{\lambda}{3!}\Phi(x){\phi}^3(x)$,
and $\hat{L}_{int}=\lint   -\tilde{L}_{int}$ .           .
The decay rate of the process (both in imaginary time formalism and in TFD)
using GCR
has been obtained   by Fujimoto et al.  \ci{fujimoto} to be
\be
\ba
\gamgcr(w)=-\dsfrac {Im \bar\Sigma(\omega)}   {\omega}=
\pi\lambda^2\ds\int\dsfrac
{ d \vec \kone d\vec \ktwo\delta(\omega-\omega_1-\omega_2-\omega_3)}
{ (2\pi)^6 8\omega\omega_1
\omega_2\omega_3}
\times
\nwl
\times
\quad
\lbr (1+n_1)(1+n_2)(1+n_3)-n_1n_2n_3\rbr
\lab{ampl4.31}
\ea
\ee
where $n_i=n_B(k_i)$, $\omega_i=\sqrt{\sqvec{k_i}+m^2} $,
 $(i=1,2,3)$ and $\omega=\sqrt{ \sqvec{k}+M^2}$.
 On the other hand the decay rate is related to the transition
 amplitude as:
\be
\ba
\Gamma(\omega)=\dsfrac{1}{2\omega}
\ds\int \dsfrac{ d\vec \kone d\vec
\ktwo d\vec \kuch (2\pi)^4\delta(k-\kone-\ktwo-\kuch)
\mod{\cmfi(T)}^2 }
{(2\omega_1)(2\pi)^3 (2\omega_2)(2\pi)^3 (2\omega_3)(2\pi)^3}
\lab{ampl4.32}
\ea
\ee
 where the amplitude $\cmfi(T)$
 is defined by Eq.\re{ampl3.4}
  with the elements
 of $\hat S$ - matrix:
\be
\matel{f}{\hat S}{i}=i\lambda
 \ds\int dx \matel{0\betaq}{b_{\kone}\betaq
  b_{\ktwo}\betaq b_{k_3}\betaq
\lbr \Phi(x)\phi^3(x)-\tilde\Phi(x){\tilde\phi}^3(x)\rbr
\akrest_k\betaq}{0\betaq}
\lab{ampl4.33}
\ee
As in the   previous example, we get
\be
\ba
\delta(\omega-\omega_1-\omega_2-\omega_3)\mod{\cmfi(T)}^2 =
\nwl
=\lbr
\ch{k}\ch{\kone}\ch{\ktwo}\ch{\kuch}-\sh{k}\sh{\kone}\sh{\ktwo}\sh{\kuch}\rbr ^2
\delta(\omega-\omega_1-\omega_2 -\omega_3)    =
\nwl
=n_1n_2n_3n_\omega
\lbr 1-\expon{ \beta(
\omega+\omega_1+\omega_2+\omega_3)/2 } \rbr  ^2
\delta(\omega-\omega_1-\omega_2-\omega_3)=
\nwl
=n_1n_2n_3n_\omega\lbr 1-\expon{\beta\omega}\rbr
\lbr 1-\expon{\beta(\omega_1+\omega_2+\omega_3         )}\rbr
\delta(\omega-\omega_1-\omega_2-\omega_3)
\nwl
=n_1n_2n_3n_\omega
\lbr 1-\dsfrac{ 1+n_\omega }{ n_\omega }\rbr
\lbr 1-\dsfrac{ (1+n_1)(1+n_2)(1+n_3) }
{n_1n_2n_3}\rbr \delta(\omega-\omega_1-\omega_2-\omega_3)=
\nwl
=\lbr (1+n_1)(1+n_2)(1+n_3)-n_1n_2n_3\rbr
\delta(\omega-\omega_1-\omega_2-\omega_3)
\ea
\lab{ampl4.34}
\ee
where $n_\omega=n_B(k)$, $n_i=n_B(k_i)$. Then,
 by using  \re{ampl4.34} in \re{ampl4.32} the result given in
   Eq.  \re{ampl4.31} is reproduced. Again the direct calculations
   and those from   GCR agree.
\section{The detailed balance.}

The last example of the previous section can be easily generalized
for the decay of a boson into $m$ particles:
 $\Phi(k)\rightarrow \phi(\kone)+\phi(\ktwo)+\dots +\phi(k_m)$
 with  the decay rate
\be
\ba
\Gamma(\omega)=\dsfrac{1}{2\omega}
\ds\int \dsfrac{ d\vec \kone d\vec \ktwo
\dots d\vec \km (2\pi)^4\delta^4(k-\kone-\ktwo-\dots -\km)
\mod{\cmfi(T)}^2 } {(2\omega_1) (2\omega_2)
 \dots (2\omega_m)(2\pi)^{3m}}
\lab{ampl5.1}
\ea
\ee
where
\be
\ba
\delta(\omega-\omega_1-\dots-\omega_m)\mod{\cmfi(T)}^2=
\delta(\omega-\omega_1-\dots -\omega_m)
\nwl
\lbr \ch{k}\ch{\kone}\dots\ch{\km}-\sh{k}\sh{\kone}
\dots\sh{\km}\rbr ^2=
\nwl
=n_1\dots n_mn_\omega\lbr 1-\expon{\beta\omega}\rbr
\lbr 1-\expon{\beta(\omega_1+\dots+\omega_m)      }\rbr
=\nwl
=\lbr (1+n_1)(1+n_2)\dots(1+n_m)-n_1n_2\dots n_m\rbr
\delta(\omega-\omega_1-\dots-\omega_m) .
\lab{ampl5.2}
\ea
\ee
It is clear
from Eq. \re{ampl5.1} and \re{ampl5.2}  that,
 the total decay rate $\Gamma(\omega) $
  is the difference of the forward rate
  of the boson decay - $\Gamma_d(\omega) $
   and the rate  of the inverse process $\Gamma_i(\omega) $:
$\Gamma(\omega) =\Gamma_d(\omega)-\Gamma_i(\omega)$, which are related
 by the principle of detailed balance:
\be
\dsfrac{\Gamma_d(\omega)}{ \Gamma_i(\omega)  }=\dsfrac{
(1+n_1)(1+n_2)\dots(1+n_m)    }{ n_1n_2\dots n_m}
=\expon{\beta(\omega_1+\omega_2+\dots+\omega_m)}=\expon{\beta\omega}
\lab{ampl5.3}
\ee
This relation was first shown by Weldon, \ci{weldon}
 by analysing the discontiniuties of the self
 energy function in the Matsubara formalism.
Thus, we conclude that, our approach of direct calculation
 of rates through elements of
  the $\hat S$ matrix leads to the correct
   relation for the detailed balance principle at  finite temperature.
   Note that, the Eqs. \re{ampl5.1} -   \re{ampl5.3}
    reveal  the importance of doubling of degrees
    of freedom at nonzero temperature: If we had neglected
     the term $\tilde\lint$ then the term proportional
     to "$\sh{k}"$ in Eq. \re{ampl5.2}
     would have vanished giving a wrong relation.
Although the "tilde" particles are introduced as an essential part
of  the finite temperature formalism
 as  fictitious particles, they play an
 important role in direct and inverse processes
  taking place at finite temperature. It is important to remark that
  there are no transitions, and hence no matrix element, between the tilde
  and non - tilde particles.

However, following Keil \ci{keil},    we underline that, $\Gamma(\omega) $
and not the partial rates $\Gamma_d(\omega)$ and $\Gamma_i(\omega)$
represent the physically measurable decay rate! The reason is that,
it is  the full decay width
$\Gamma(\omega) =\Gamma_d(\omega)-\Gamma_i(\omega)$ that is
  connected to the pole of the Green's function \ci{fujimoto, saito}
\be
\ba
\Delta_{11}(k)=\dsfrac{1+n_B(k_0)} { k^2-m_{0}^{2}-\bar\Sigma+i\varepsilon}-
\dsfrac{n_B(k_0)} { k^2-m_{0}^{2}-\bar\Sigma^*-i\varepsilon}=
\nwl
\nwl
\dsfrac{1+n_B(k_0)} { (k_0+i\Gamma/2)^2-\omega^2 +i\varepsilon}-
\dsfrac{n_B(k_0)}{ (k_0-i\Gamma/2)^2-\omega^2 -i\varepsilon}
\lab{ampl5.4}
\ea
\ee
where $\omega^2=k_{0}^{2}+Re\bar\Sigma-\Gamma^2/4$,
 $\Gamma=-Im\bar\Sigma/k_0$, $k_0=\sqrt{\sqvec{k}+m^2}$.
\section{Scattering cross section of $1+2\rightarrow \pr{1}+\pr{2} $ .}
\indent
As it was stated in Sect. III  in our approach, in accordance with the
concept of TFD, the relation between the rate
 (cross section) and the transition amplitude is the
  same as it is in the zero temperature quantum  field theory.
   For example the cross section of the process
$a(\kone)+b(\ktwo)\rightarrow
\pr{a}(\pkone)+\pr{b}(\pktwo)$
is \ci{mandlshaw}
\be
\dsfrac{d\sigma}{d\pr{\Omega}}=\dsfrac{ \mod{\cmfi(T)}^2
(\prod_{l}(2m_l))\mod{\vec \pkone}^2 }
{ 64\pi^2v_{\mbox{rel}}\omega_1 \omega_2\pom_1\pom_2 }
\lbrace\dsfrac{ \partial (\pom_1+\pom_2)} {\partial \mod{\vec \pkone}
}\rbrace^{-1}
\lab{ampl6.1}
\ee
in the usual notation.
In particular, the cross section of the elastic scattering
of two bosons in their center of mass (CoM)
 frame ($\vec \pkone=-\vec\pktwo)$ is
\be
(\dsfrac{d\sigma}{d\Omega})=\dsfrac{ \mod{\cmfi(T)}^2}
{64\pi^2(\omega_1+\omega_2)^2}
\lab{ampl6.2}
\ee
These equations imply that
\be
\dsfrac{ (d\sigma/d\pr{\Omega})|_{T\neq 0}   }
{ (d\sigma/d\pr{\Omega})|_{T=0}   }=
\dsfrac{ \mod{\cmfi(T\neq0)}^2  }{ \mod{\cmfi(T=0)}^2 }\equiv W(T)
\lab{ampl6.3}
\ee
where the amplitudes may be  determined through  Eq.s
\re{ampl3.2}-\re{ampl3.4}.
To see  the consequences of this relation we will consider some examples.
\nwl
{\bf A.  Boson - Boson scattering.}

Let us assume that both a and b particles are
 bosons with the interaction Lagrangian
$\lint=\lambda \phi_{a}^{2}(x)\phi_{b}^{2}(x)$
and $\hat{L}_{int}=\lint   -\tilde{L}_{int}$ .
At  the tree level, using  the results given in
Eq.s \re{ampl3.2} - \re{ampl3.4} and Eq. \re{ampl4.11}
 in \re{ampl6.3}
we get
\be
\ba
W_{BB}(T)=\dsfrac{\mod{\cmfi(T\neq0)}^2 }{\mod{\cmfi(T=0)}^2 }=
\lbr C(T)-S(T)\rbr ^2=
\nwl
=n_1n_2\prn_1\prn_2
\lbr \expon{\beta(\omega_1+\omega_2+\pom_1+\pom_2)/2)}-1\rbr ^2,
\lab{ampl6.4}
\ea
\ee
where
\be
\ba
C(T)=\ch{\kone}\ch{\ktwo} \ch{\pkone}\ch{\pktwo},
\nwl
S(T)=\sh{\kone}\sh{\ktwo} \sh{\pkone}\sh{\pktwo},
\lab{ampl6.5}
\ea
\ee
and $n_i=n_B(k_i)$, $\prn_i=n_B(\pr{k}_i)$.
 Due to  energy conservation we have
\be
\lbr \expon{\beta(\omega_1+\omega_2+\pom_1+\pom_2)/2}-1\rbr ^2
=(\expon{\beta(\omega_1+\omega_2)}-1)(\expon{\beta(\pom_1+\pom_2)}-1).
\lab{ampl6.6}
\ee
Now we may express the exponents via $n_i$, $\prn_i$
\be
\ba
\expon{\beta\omega_i}=\dsfrac{1+n_i}{n_i},
\quad
\quad
\expon{\beta\pom_i}=\dsfrac{1+\prn_i}{\prn_i};
\lab{ampl6.7}
\ea
\ee
to get
\be
\ba
W_{BB}(T)=(1+n_1+n_2)(1+\prn_1+\prn_2)=
\nwl
=\lbr (1+n_1)(1+n_2)-n_1n_2\rbr \lbr (1+\prn_1)(1+\prn_2)-\prn_1\prn_2\rbr .
\lab{ampl6.8}
\ea
\ee
\nwl
{\bf B. Fermion - Fermion scattering.}

Now let us assume that both $a$ and $b$
 particles are  fermions
with the interaction lagrangian
$\lint=g\bar\psi_a(x)\Gamma_\alpha\psi_a(x)\bar\psi_b(x)\Gamma^\alpha\psi_b(
x)$
where $\Gamma_\alpha$  is an appropriate combination of Dirac matrices.
Similarly as in the  previous case we get :
\be
\ba
W_{FF}(T)=\dsfrac{\mod{\cmfi(T\neq0)}^2 }{\mod{\cmfi(T=0)}^2 }
=\lbr \cs{\kone}\cs{\ktwo} \cs{\pkone}\cs{\pktwo}  -
\sn{\kone}\sn{\ktwo} \sn{\pkone}\sn{\pktwo}\rbr ^2=
\nwl
=n_F(\kone)n_F(\ktwo)n_F(\pkone)n_F(\pktwo)
\lbr \expon{\beta(\omega_1+\omega_2+\pom_1+\pom_2-\mu_1-
\mu_2-\pr{\mu}_1-\pr{\mu}_2)/2}  -1\rbr ^2.
\lab{ampl6.9}
\ea
\ee
At equilibrum $ \mu_1+\mu_2=\pr{\mu}_1+\pr{\mu}_2 $,
and hence, by using
\be
\ba
\lbr \expon{\beta(\omega_1-\mu_1+\omega_2-\mu_2)}-1\rbr ^2=
\lbr \expon{\beta(\omega_1-\mu_1+\omega_2-\mu_2)}-1\rbr
\lbr \expon{\beta(\pom_1-\pr{\mu}_1+\pom_2-\pr{\mu}_2)}-1\rbr =
\nwl
=\lbr \dsfrac{1}{ n_F(\kone)n_F(\ktwo) } - \dsfrac{1}{ n_F(\kone) }
-\dsfrac{1}{ n_F(\ktwo)}\rbr
\lbr \dsfrac{1}{ n_F(\pkone)n_F(\pktwo) }
- \dsfrac{1}{ n_F(\pkone) }-\dsfrac{1}{ n_F(\pktwo)}\rbr ,
\lab{ampl6.10}
\ea
\ee
we  get
\be
\ba
W_{FF}(T)=\dsfrac{\mod{\cmfi(T\neq0)}^2 }{\mod{\cmfi(T=0)}^2 }
=\lbr (1-n_1)(1-n_2)-n_1n_2\rbr \lbr (1-\prn_1)(1-\prn_2)-\prn_1\prn_2\rbr .
\lab{ampl6.11}
\ea
\ee
where  $n_i\equiv n_F(k_i)$, and $\prn_i\equiv n_F(\pr{ k_i})$.
\nwl
{\bf C.  Fermion - Boson scattering.}

Let us assume that particle $a$  is a fermion
and  particle $b$ is a
boson with the interaction lagrangian given as
 $\lint=-ig\bar\psi(x)\Gamma\psi(x)\phi(x) $
 ( for example $\gamma=\gamma_5$ for pion - nucleon scattering.)
In this case we get at  tree level
\be
\ba
W_{FB}(T)=\dsfrac{\mod{\cmfi(T\neq0)}^2 }{\mod{\cmfi(T=0)}^2 }
=\lbr \cs{\kone}\ch{\ktwo} \cs{\pkone}\ch{\pktwo}  +
\sn{\kone}\sh{\ktwo} \sn{\pkone}\sh{\pktwo}\rbr ^2 .
\lab{ampl6.12}
\ea
\ee
Then using  similar manipulations,
 as in previous examples we  get
\be
\ba
W_{FB}(T)=\dsfrac{\mod{\cmfi(T\neq0)}^2 }{\mod{\cmfi(T=0)}^2 }
=\lbr 1-n_F(\kone)+n_B(\ktwo)\rbr \lbr 1-n_F(\pkone)+n_B(\pktwo)\rbr .
\lab{ampl6.13}
\ea
\ee

The expressions \re{ampl6.8}, \re{ampl6.11} and \re{ampl6.13} for the
relation of in medium cross sections of
$1+2\rightarrow 1^\prime+2^\prime$ scattering can
be interpreted physically in the spirit of Weldon \ci{weldon}.
For example    $W_{FB}(T)$   can be rewritten as
\be
\ba
W_{FB}(T)=(1-n_F(\kone))(1+n_B(\ktwo))(1-n_F(\pkone))(1+n_B(\pktwo))+
\nwl
+n_F(\kone)n_B(\ktwo)n_F(\pkone)n_B(\pktwo)+
n_F(\kone)n_B(\ktwo)(1-n_F(\pkone))(1+n_B(\pktwo))+
\nwl
+n_F(\pkone)n_B(\pktwo)(1-n_F(\kone))(1+n_B(\ktwo)) .
\lab{ampl6.14}
\ea
\ee
It is to be emphasized that the in-medium cross sections at finite
temperature are not directly measureable in any many-body process.
However such cross-sections are an important input to treat a gas at
low density through a kinetic theory approach such as a Boltzmann
equation. The finite temperature cross-sections will be a part of the
collision term that would lead to a distribution function by a
solution of the Boltzmann equation. As indicated in the introduction
these types of results are a necessary input to a better
understanding of both the early Universe and the hadronic matter at
finite temperature and density.

\section {Loop corrections}
All the above results have been obtained at the tree level. Now we wish
to show  how the method can be handled beyond
the tree level. For simplicity
we choose the $\phi^4(x)$ interaction
:$\lint=-\dsfrac{\lambda}{4!}: \phi^4(x):$  .
The $\hat S$ matrix in Eq. \re{ampl2.12} up to order $\lambda^2$ is given
by:
\be
\ba
\hat S\approx 1-i\ds\int dx \hat H_I(x) +
\dsfrac{(-i)^2   }{2!}
\ds\int dx dy
T\lbr \hat H_I(x)\hat H_I(y)\rbr
 \equiv \hat{S}^{(1)}+\hat{S}^{(2)}.
\lab{ampl7.1}
\ea
\ee
For the scattering process from the initial state
\be
\ket {i} =\ket{\phi(\kone) \phi(\ktwo)}=
 \akrest_{k_1}\betaq\akrest_{k_2}\betaq
 \ket{0(\beta)}
 \lab{ampl7.2}
 \ee
 to a final state
  \be
\ket {f} =\ket{\phi(\pkone) \phi(\pktwo)}=
 \akrest_{\pkone}\betaq\akrest_{ \pktwo}\betaq
 \ket{0(\beta)}
 \lab{ampl7.3}
 \ee
 the matrix element  $\matel{f}{ \hat{S}^{(1)}}{i}$ and hence
 the amplitude $\cmfi(T)=\cmfione(T)+\cmfitwo(T)       $,
  is defined as,
 \be
\matel{f}{\hat S^{(1)}+  \hat S^{(2)}   }{i}=
\delta_{if}+\delta^4(P-P^\prime)
\dsfrac{\cmfione(T)+\cmfitwo(T)      }
{ (4\pi)^2\sqrt{\omega_1\omega_2\pom_1\pom_2}     },
\lab{ampl7.4}
\ee
with $P=k_1+k_2$  and $P^\prime=\pkone+\pktwo$.
$\cmfione(T)$ and $\cmfitwo(T)$
 can be simply calculated by using the procedure used in sec. VI.
 For instance,
 \be
 \cmfione(T)=-i\lambda\lbr C(T)-S(T)\rbr
 \lab{ampl7.5}
 \ee
 where $C(T)$ and $S(T)$ are defined  by  Eqs \re{ampl6.5}.

To evaluate  $\cmfitwo(T)$,
the generalized Wick theorem \ci{ojima} is used and
 $2\times2$ Green's function $\Delta_{ab}(q)$ (see Appendix) is
introduced. In particular:
 \be
 \ba
 i\Delta_{11}^{0}(x-y)=\bra{0\betaq} T\lbr \phi(x)\phi(y)\rbr
 \ket{0(\beta)} =
 i\ds\int
 \frac{d^4q}{(2\pi)^4} \expon{ -iq(x-y)} \Delta_{11}^{0}(q)
 \nwl
 i\Delta_{22}^{0}(x-y)=\bra{0\betaq}
 T\lbr\tilde\phi(x)\tilde\phi(y)\rbr \ket{0(\beta)}=i\ds\int
 \frac{d^4q}{(2\pi)^4} \expon{ -iq(x-y)} \Delta_{22}^{0}(q)
 \nwl
 i\Delta_{12}^{0}(x-y)=\bra{0\betaq}
 T\lbr\phi(x)\tilde\phi(y)\rbr \ket{0(\beta)}=i\ds\int
 \frac{d^4q}{(2\pi)^4} \expon{ -iq(x-y)} \Delta_{12}^{0}(q) .
 \lab{ampl7.6}
\ea
\ee
In terms of these Green's functions, the amplitude is proportional to
$\lambda^2$ and is
\be
\ba
 \cmfitwo(T)=
 \lambda^2\{ C(T)\ds\int\frac{dq^4}{ (2\pi)^4 }
 \Delta_{11}^{0}(q)\Delta_{11}^{0}(P-q)
 +S(T)\ds\int\frac{dq^4}{ (2\pi)^4
 }\Delta_{22}^{0}(q)\Delta_{22}^{0}(P-q)-
\nwl
- \ds\int\frac{dq^4}{ (2\pi)^4 }
 \lbr C_s(k_1,k_2;\pkone,\pktwo)\Delta_{12}^{0}(q)\Delta_{12}^{0}(P-q)+
 C_s(\pkone,\pktwo;k_1,k_2)\Delta_{21}^{0}(q)\Delta_{21}^{0}(P-q)
 \rbr\} .
 \lab{ampl7.7}
\ea
 \ee
where $C_s(k_1,k_2;\pkone,\pktwo)=
\ch{\kone}\ch{\ktwo}\sh{\pkone}\sh{\pktwo}$.
 The  terms in the integrand are related by
 \be
\ba
 \Delta_{11}^{0}(q)=-\lbr \Delta_{22}^{0}(q)  \rbr ^*=
 \dsfrac{1}{q^2-m^2+i\varepsilon}-
 2i\pi n_B(q)\delta(q^2-m^2),
\nwl
\Delta_{12}^{0}(q)=\Delta_{21}^{0}(q)
 \lab{ampl7.8}
\ea
 \ee
 and hence  Eq. \re{ampl7.7} simplifies to
 \be
 \ba
 \cmfitwo(T)=\lambda^2\{C(T)F(P,T)+S(T)F^*(P,T)+
\nwl
+
8\pi^2e^{\beta P_0/2}I_3(P)\lbr C_s(k_1,k_2;\pkone,\pktwo)+
C_s(\pkone,\pktwo;k_1,k_2)\rbr\},
 \nwl
 F(P,T)=\ds\int\frac{dq^4}{ (2\pi^4)}
 \Delta_{11}^{0}(q)\Delta_{11}^{0}(P-q)=
 I_1(P)-4\pi i I_2(P)-4\pi^2 I_3(P)
 \lab{ampl7.9}
 \ea
 \ee
with
\be
\ba
I_1(P)=\ds\int \dsfrac{dq^4}{ (2\pi)^4
}\dsfrac{1}{(q^2-m^2+i\varepsilon) ((P-q)^2-m^2+i\varepsilon) }
\nwl
\nwl
I_2(P)=\ds\int \dsfrac{dq^4}{ (2\pi)^4 }\dsfrac{ n_B(q)\delta(q^2-m^2)
}{ (P-q)^2-m^2 }
\nwl
\nwl
I_3(P)=\ds\int \dsfrac{dq^4}{ (2\pi)^4 }
n_B(q)n_B(P-q)\delta(q^2-m^2) \delta ((P-q)^2-m^2) .
\lab{ampl7.10}
\ea
\ee
The integrals $I_1(P), I_2(P), I_3(P)$
can be evaluated in the CoM frame of the two  particles , where
$P^2=(\kone^0+\ktwo^0)^2=E^2$ - total energy of the system.

The integral $I_1(P)$ is logarithmically divergent
and can be treated  using a renormalization procedure
 and this finite contribution
is called the vacuum fluctuation  correction. This term
  exists even at zero temperature, and  can be
calculated by the  usual method of dimensional regularization \ci{ramond}.
As a result   $I_1(P)$ may be written  as:
\be
\ba
I_1(P)=\dsfrac{iN_\varepsilon}{16\pi^2} +
\dsfrac{i}{16\pi^2}\lbr b_0(E)-\lnmu\rbr -\dsfrac{\sqrtd}{16\pi}
\nwl
b_0(E)=2+ \sqrtd \ln \mod{\dsfrac{ 1-\sqrtd}{ 1+\sqrtd}}
\nwl
D_0=1-4m^2/E,
\lab{ampl7.11}
\ea
\ee
where the first term  involving
$N_\varepsilon=1/\varepsilon-\gamma_0+\ln 4\pi$ is infinite. To
eliminate the latter we introduce the following counterterm in
the Lagrangian \ci{ramond}
\be
L_{\mbox{contr}}=-\dsfrac{\lambda^2 N_\varepsilon }{ 16\pi^24!}
:\phi^4(x):
\lab{ampl7.12}
\ee
which clearly does not depend on temperature. On the other hand, in
accordance with the tilde oparation rule, this automatically leads us to
introduce its tilde partner also:
\be
\tilde L_{\mbox{contr}}=
-\dsfrac{\lambda^2 N_\varepsilon} {
16\pi^24!}:\tilde\phi^4(x):
\lab{ampl7.13}
\ee
which serves to
compensate the divergence, coming from
$S(T)*F^*(P,T)$ term in  Eq. \re{ampl7.9}.

The $I_2(P)$ is convergent and can  be evaluated numerically:
\be
\ba
I_2(P)=\ds\int \dsfrac{ d \vec q }{ (2\pi)^4 } \ds\int_{-\infty}^{+\infty}
\dsfrac{ d q_0 n_B(q_0)
\lbr \delta(q_0+\mod{\omega_q})-\delta(q_0-\mod{\omega_q})\rbr  }
{ 2\omega_q ((P-q)^2-m^2) }=
\nwl
\nwl
=\dsfrac{1}{4\pi^3}\ds\int
\dsfrac{dq q^2 n_B(q)}
{ E^2 (1-4\omega_{q}^{2}/E^2) \omega_q  }\equiv\dsfrac{1}{64\pi^3}\bar
I_2(E,T)
\lab{ampl7.14}
\ea
\ee

The integral $I_3$ including two $ \delta-$ functions may be evaluated
in a similar way as in sect. IV (Eq.\re{ampl4.6})  and is given as
\be
\ba
I_3(P)=\ds\int \dsfrac{dq_0 d\vec q n_B(P-q)n_B(q)}{ (2\pi)^4 4E\omega_q}
 \delta(q_0-\omega_q)\delta(\omega_q-E/2)=
 \dsfrac{n_{B}^{2}(E/2)\sqrtd} { 32\pi^3} .
 \lab{ampl7.15}
 \ea
 \ee

 Now , using Eqs.  \re{ampl7.5}, \re{ampl7.9}-\re{ampl7.15} and
 introducing $\cmfibar=i\cmfi $
  gives
  \be
  \ba
  \cmfibar(T)=Re \cmfibar(T)  +  iIm \cmfibar(T)
  \nwl
  Re \cmfibar(T)=\lambda(C(T)-S(T))(1-\bar\lambda A(E,T))
  \nwl
  Im \cmfibar(T)=-\lambda^2\sqrtd \lbr(C(T)+S(T))
 (1+2n_{B}^{2})-4C_s(T)e^{\beta E/2}n_{B}^{2}
 \rbr/(16\pi),
  \lab{ampl7.16}
  \ea
  \ee
  where $A(E,T)=b_0(E)-\lnmu-\bar I_2(E,T)$, $n_{B}=n_B(E/2)$,
  $C(T)=(1+n_B)^2$ and $S(T)=n_{B}^{2}$, $C_s(T)=2e^{\beta E/2}n_{B}$ and
$\bar\lambda=\lambda/16\pi^2$

  From \re{ampl7.16} one may get the relation for the in -  medium elastic
  cross section in $\phi^4(x)$ scalar  theory up to one loop approximation:
 \be
 \ba
\dsfrac{ (d\sigma/d\pr{\Omega})|_{T\neq 0}   }
{ (d\sigma/d\pr{\Omega})|_{T=0}   }=
\dsfrac{ (C(T)-S(T))^2(1-\bar\lambda A(E,T))^2+
\pi^2{\bar\lambda}^2D_0
\lbr (1+2n_{B}^{2})(C(T)+S(T))-8e^{\beta E}n_{B}^{3}\rbr^2}
{ \lbr 1-\bar\lambda(b_0(E)-\lnmu)\rbr ^2+\pi^2 D_0 {\bar\lambda}^2 }
  \lab{ampl7.17}
  \ea
  \ee

The appropriate Feynman rules for $\phi^4$  scalar theory at finite
temperature can be explicitly obtained  from Fig. 2.
They are similar to those for zero temperature. The main differences
are the following.\\
 The amplitude of any order $n$
consists of $n$ vertices. There are two kinds of vertices: "dotted"
(ordinary)
and "circled" ones.
To calculate $\cmfi(T)$ \\
1) Attach factor $-i\lambda$ to each "dotted" vertex.\\
2) Attach the factor $\ch{p}$ to each external
leg (incoming or outgoing particle) with momentum $p$
connected to the "dotted"
vertex.\\
3) Attach the factor $\sh{p}$ to each external
leg (incoming or outgoing particle) with momentum $p$
connected to the "circled "
vertex.\\
4) Attach $i\Delta_{11} (q)$ to the propagator carrying momentum $q$,
 connecting two "dotted" vertices.\\
5) Attach $i\Delta_{22} (q)$ to the propagator carrying momentum $q$,
 connecting two "circled" vertices.\\
6) Attach $i\Delta_{12} (q)$ to the propagator carrying momentum $q$,
 between a "dotted" and a  "circled" vertices.\\
7) Attach $i\Delta_{21} (q)$ to the propagator carrying momentum $q$,
 between a   "circled" and a "dotted" vertices.\\
Using these rules one may calculate any process in $\phi^4$
theory at finite temperature to  any order of perturbative theory.

{\bf Gluon Self-energy and Screening length.}

The gluon self-energy in one looop order is calculated in order to
obtain the screening length in a quark-gluon plasma. The calculation
is straight forward in Thermo Field Dynamics. Working in Feynman gauge
the ghost states do not contribute. Then the photon self-energy
$\pi(\vec{k}, k_{0}; T)$ in the limit $\vec{k} \rightarrow 0$ and
$k_{0} \rightarrow 0$ but at finite temperature reduces to a simple
form. The screening length is related to real part of the self energy
which has the expression in one loop order as
$$
Re\pi^{ab}_{00} (\vec{k} \rightarrow 0, k_{0} \rightarrow 0 ; T \pm
0) = \frac{2 g^{2} T^{2} \delta^{ab}}{\pi^{2}} (N + \frac{1}{2}
N_{f}) \Gamma(2) \xi(2) = \delta^{ab} \frac{1}{3}(N + \frac{N_{f}}{2})
g^{2} T^{2}
$$

where $N_{f} =$ number of flavors and a, b refer to the color indices
of the group SU(N). Here the term proportional to N arises from the
triplet and quartet couplings of gluons and the term proportional to
$N_{f}$ arises from quark-anti-quark loops. Then the color electric
mass of the gluon is
$$
m^{2}_{e \ell} = \frac{1}{3} (N + \frac{1}{2} N_{f}) g^{2}T^{2}
$$
and the screening length at finite temperature is
$$
L(T) = \frac{\sqrt{3}}{\sqrt{N + \frac{1}{2} N_{f} gT}}
$$
The color magnetic mass of a gluon at finite temperature is zero.
Details of this calculation are given by Zhao and Khanna [23].

\section{Discussions and summary}
In the present paper
the decay amplitudes and scattering processes are studied in TFD.
We have proved that, in the TFD formalism,
 the rate and transition amplitude  of a process can
 be evaluated in a similar way as in the zero temperature case.
Note that, both imaginary and real parts the amplitudes at finite
temperature can be calculated directly.
The amplitudes are written by establishing a set of Feynman rules using
the Lagrangian $\hat L=L-\tilde L$. The Feynman diagrams are
considered. Use of Bogoluibov transformations to define a pure vacuum
state leads to Feynman rules for including temperature dependant
factors for a process occuring in a system at finite temperature. It
will be interesting to evaluate for example $NN$ scattering in
nuclear matter at finite temperature. Temperature dependence of
coupling constants and masses  have been explored in  perturbative
theory  of TFD \ci{ourtfd, songgao}.
 At present we are attempting a self consistent calculation of
these quantities.
The role of Ward Takahashi relations at finite T will be
crucial to obtaining  a self consistent solution.

The present discussion for nuclear matter in equilibrium at finite T
can be easily extended to QGP at temperature above the deconfinement
phase transition. Some of the  interesting properties like screening
length and quark - quark interaction at finite temperature have been
investigated at finite T \ci{zhaokhanna}.
 The results are obtained in a simple and
straightforward manner. There  are numerous other properties of nuclear
matter and QGP that may be calculated  and would be helpful in studies of
many particle hadronic systems.

The method developed in this paper using Thermo Field Dynamics are
applicable to calculations at tree level as well as loop calculations.
This will allow us to calculate processes in nuclear matter as well as
in quark gluon plasma at finite temperature.

The present approach  simplifies  calculation of
 concrete processes taking place in hot and dense matter.
 Examples are neutrino electron scattering in thermal
 plasma \ci{hardy, becs} , $ \bf\beta$ - decay, which plays
  an important role in cosmology, in -medium nucleon -
   nucleon cross sections \ci{machelidt}
taking place in ion - ion collisions etc.
These processes are being  investigated with the present approach.

\section*{Acknowledgments}
We thank  A.E. Santana, M. Revzen and R. Kobes   for useful discussions.
  A.M. Rakhimov            is
indebted to the  University of Alberta for hospitality
 during his stay, where   the main part of
this work was  performed. The research of F.C. Khanna is
 supported in part by
Natural  Sciences and Engineering Research Council of Canada. We thank
the referee for some important comments and for pointing out a serious
mistake.

\newpage
\appendix
\section{}
\label{AP.A}
{\bf A. The nessessity of doubling.}

In the statistical mechanics the statistical average of a quantity $
A$ is given by
\be
<A>=Z^{-1}\betaq\Tr\lbr A\expon{-\beta\cal{H} }\rbr
\lab{amplA1}
\ee
where ${\cal H}=H-\mu N$ , $Z\betaq= \Tr\lbr \expon{-\beta\cal{H} }\rbr $.
In the zero temperature field theory the average of an observable $A$
is determined by it's vaccuum expectation value:
\be
<A>=\matel{0}{ A}{0} .
\lab{amplA2}
\ee
To make the temperature formalism completely parallel to the zero
temperature case we assume that there exists a " thermal vaccuum
state",  such that
\be
<A>=\matel{0(\beta)}{ A}{0(\beta)}=
Z^{-1}\ds\sum_n\expon{-\beta E_n}\matel{n}{A}{n}
\lab{amplA3}
\ee
for any operator $A$. Since the Hilbert space is complete we can
expand $\tvacr$ in terms of $\ket{n}$ as
\be
\tvacr=\ds\sum_n \ket{n}\braket{n}{0\betaq}\equiv\ds\sum_n f_n\betaq\ket{n} .
\lab{amplA4}
\ee
>From the last equation we obtain
\be
\matel{0(\beta)}{ A}{0(\beta)}=
\ds\sum_{n,m}f_{n}^{*}\betaq f_{m}\betaq    \matel{n}{A}{m}
\lab{amplA5}
\ee
which will agree with eq. \re{amplA3} provided
\be
f_{n}^{*}\betaq f_{m}\betaq=Z^{-1}\expon{-\beta E_n}\delta_{mn} .
\lab{amplA6}
\ee
However as we see from eq. \re{amplA4}, $f_n\betaq$'s are ordinary
numbers and therefore it is not possible to satisfy \re{amplA6}.
This shows that as long as we restrict ourselves to the original
Hilbert space, we cannot define a finite temperature vacuum which
would satisfy eq. \re{amplA3}. Therefore we have to double the Hilbert
space, introducing a fictitious system identical to the original
system. Let us denote this auxiliary system as a tilde system.
Mathematically this means that
\be
\ket{n,\tilde m}=\ket{n}\otimes\ket{\tilde m} .
\lab{amplA7}
\ee
Now we can expand $\tvacr$ in terms of these $\ket{n,\tilde m}$
states
\be
\tvacr=\ds\sum_n f_n(\beta)\ket{n,\tilde n}=
\ds\sum_n f_n(\beta)\ket{n}\otimes\ket{\tilde n} .
\lab{amplA8}
\ee
In this case we note that
\be
\ba
\matel{0(\beta)}{ A}{0(\beta)}=
\ds\sum_{n,m}f_{n}^{*}\betaq f_{m}\betaq\matel{n,\tilde
n}{A}{m,\tilde m}=
\nwl
=\ds\sum_{n,m}f_{n}^{*}\betaq f_{m}\betaq
\matel{n}{A}{m}\delta_{nm}=\ds\sum_{n}f_{n}^{*}\betaq f_{n}\betaq
\matel{n}{A}{n} .
\lab{amplA9}
\ea
\ee
Here we have used the fact that an operator of the physical system does not
act on
the Hilbert space of the tilde system and vice versa. So,
using the orthonormality of the states we can write
\be
\ba
\matel{n,\tilde m}{A}{n^\prime,{\tilde m}^\prime   }=
\matel{n}{A}{n^\prime}\braket{\tilde{m}}{ { \tilde{\pr{m}}}
 }=
\delta_{m m^\prime}\matel{n}{A}{n^\prime}
\nwl
\matel{n,\tilde m}{\tilde A}{n^\prime,{\pr{\tilde m}}  }=
\matel{n}{\tilde A}{n^\prime}
\braket{\tilde{m}}{ {\pr{\tilde{m}}} }
=
\delta_{m m^\prime}\matel{n}{\tilde A}{n^\prime} .
\lab{amplA10}
\ea
\ee
>From \re{amplA3} and  \re{amplA6} it is seen that
\be
f_{n}^{*}\betaq =f_{n}\betaq=Z^{-1/2}(\beta)\exp(-\beta E_n/2)
\lab{amplA11}
\ee

Now we can write the Fock space based on the vacuum $\tvacr$ which is
defined by a Bogoliubov transformation as
\be
\tvacr=U\betaq\ket{0,0}
\lab{amplA12}
\ee
where
\be
U\betaq=\exp\lbr -\theta(a\tilde a-\akrest\takrest)\rbr
\lab{amplA13}
\ee
where $\cosh(\theta)=1/\sqrt{(1-\exp(-\beta\omega))} $ and $ \omega$
is the energy.
Then the annihilation operators at finite temperature are defined as
\be
\ba
a\betaq=U^{-1}\betaq a U\betaq , \quad \quad
\tilde a\betaq=U^{-1}\betaq \tilde a U\betaq
\lab{amplA14}
\ea
\ee
such that
\be
\ba
a\betaq\tvacr=0 \quad, \quad   \tilde a\betaq\tvacr=0 .
\lab{amplA15}
\ea
\ee

Then Fock space is given  as
\be
\ba
\ket{0\betaq}, \quad \akrest(\beta)\ket{0\betaq}, \quad \takrest
\betaq\ket{0\betaq}, \dots
\nwl
\dots \quad \quad \frac{1}{\sqrt{n!}}
\frac{1}{\sqrt{m!}}\lbr \akrest(\beta)\rbr ^n
\lbr \takrest(\beta)\rbr ^m \ket{0\betaq}
\lab{amplA16}
\ea
\ee
where $n$ and $m$ are the number of physical particles
and tilde  particles respectively.
This therefore shows that we can introduce a temperature dependent
vacuum state so that the statistical ensemble average of any operator
can be identified with the expectation value of the operator in this
state. This, however entails a doubling of the Hilbert space.
The advantage, on the other hand, lies in the fact that the
techniques of zero temperature field theory can now be carried over to
the finite temperature case.


Here are the tilde conjugation rules \ci{umezawa}
\be
\ba
\widetilde{(AB)}=\tilde A \tilde B,
\nwl
(c_1A+c_2B\widetilde{)}=c_{1}^{\ast}\tilde A+c_{2}^*\tilde B,
\nwl
\tilde{(A^\dagger)}=(\tilde A)^\dagger,
\nwl
\widetilde{\tvacr}=\tvacr .
\lab{amplA17}
\ea
\ee
 For example the tilde partner of the weak interaction
 Lagrangian
 \be
 { L}_{W}=\dsfrac{G_F}{\sqrt{2}}\bar\nu_\mu\Gamma_\alpha
 \mu\bar{ e}
 \Gamma^\alpha\nu_e
 \lab{amplA18}
 \ee
  (in usual notations) is
\be
  \tilde{L}_{W}=\dsfrac{G_F}{\sqrt{2}}
  \tilde{\bar\nu}_e\tilde{\Gamma}_\alpha
  \tilde e\tilde{\bar\mu} \tilde {\Gamma}^\alpha\tilde\nu_\mu,
\lab{amplA19}
\ee
  but not
  $\dsfrac{G_F}{\sqrt{2}}\tilde{\bar\nu}_\mu\tilde
 {\Gamma}_\alpha\tilde{\mu}\tilde{\bar e}
  \tilde{\Gamma}^\alpha\tilde{\nu}_e$
 which would give zero contribution to the
 amplitude of the process
  $\mu\rightarrow e\bar{\nu}_e\nu_\mu $ at tree level .\\

{\bf B. Green's functions.}

In TFD the thermal doublet for each field is defined as
\begin{equation}
\phi^{(a)} (x) \equiv
\left\{
\begin{array}{c}
\phi (x)\\
 \  \tilde{\phi}(x)
\end{array}
\right\}
\equiv
\left\{
\begin{array}{c}
\phi_1 (x)\\
  \phi_2(x)
\end{array}
\right\}
\lab{amplA20}
\end{equation}
\begin{equation}
\psi^{(a)} (x) \equiv
\left\{
\begin{array}{c}
\psi (x)\\
i \  ^t\tilde{\psi}^\dagger(x)
\end{array}
\right\}
\equiv
\left\{
\begin{array}{c}
\psi_1 (x)\\
  \psi_2(x)
\end{array}
\right\}
\lab{amplA21}
\end{equation}
where $t$ means transposition with respect the spinor index,
and $a (=1,2) $ specifies a component of the  thermal doublet. The first
component is physical, and the second  is fictitious.

The scalar
 free
field may be written as  \ci{saito}:
\be
\ba
\phi(x)=\dsfrac {1} { (2\pi)^{3/2} } \ds\int \dsfrac{ d\vec p }{
\sqrt {2\omega_p}   }\lbr a_p\expon{-ipx}+\akrest_p\expon{ipx}\rbr ,
\nwl
\tilde\phi(x)=\dsfrac{ 1 } { (2\pi)^{3/2} }\ds\int \dsfrac{ d\vec p }{
\sqrt{2\omega_p} }\lbr \ta_p\expon{ipx}+\takrest_p\expon{-ipx}\rbr .
\lab{amplA22}
\ea
\ee
A  Fermion field may be written as :
\be
\ba
\psi(x)=\summa_{s}\ds\int {d\vec p}N_p\lbr
c_{p,s}u(p,s)\expon{-ipx}+\dkrest_{p,s}v(p,s)\expon{ipx}\rbr ,
\nwl
\tilde\psi(x)=\summa_{s}\ds\int {d\vec p}N_p\lbr \tc_{p,s}\tilde
u(p,s)\expon{ipx}+\tdkrest_{p,s}\tilde v(p,s)\expon{-ipx}\rbr ,
\nwl
\bar\psi(x)=\summa_{s}\ds\int {d\vec p}N_p\lbr \ckrest_{p,s}\bar u(p,s)
\expon{ipx}+d_{p,s}\bar v(p,s)\expon{-ipx}\rbr ,
\nwl
\bar {\tilde\psi} (x)=\summa_{s}\ds\int {d\vec p}N_p\lbr \tckrest_{p,s}\bar
{\tilde u} (p,s)\expon{-ipx}+\td_{p,s}\bar {\tilde v}(p,s)\expon{ipx}\rbr ,
\lab{amplA23}
\ea
\ee
where $N_p=\sqrt{ m/ (2\pi)^{3} E_{p} }$ ,
 $E_{p}=\sqrt{  m^2+\sqvec{p} }$,
 $\tilde u(p,s)=u^{*}(p,s)$ etc.

The boson  and fermion  propagators,
$\Delta_{ab}(k)$  and  $S_{ab}(k)$ respectively,  are
$2\otimes2$ matrices. They are
defined as:
\be
 \ba
 i\Delta_{ab}(x-y)=\bra{0\betaq} T\lbr \phi_a(x)\phi_b(y)\rbr \ket{0(\beta)}
 =i\ds\int
 \frac{dp^4}{(2\pi^4)} \expon{ -iq(x-y)} \Delta_{ab}^{0}(q)
 \nwl
 iS_{ab}(x-y)=\bra{0\betaq} T\lbr \psi_a(x)\bar\psi_b(y)\rbr \ket{0(\beta)}
 =i\ds\int
 \frac{dp^4}{(2\pi^4)} \expon{ -iq(x-y)} S_{ab}^{0}(q)
 \lab{amplA24}
\ea
\ee
The free propagators may be presented as a sum of Feynman propagator,
$\Delta_{F}^{0}(k)$  $\quad (S_{F}^{0}(k))$ and temperature dependent
 part, $\Delta_{T}^{0}(k)$,  $\quad (S_{T}^{0}(k))$
as:
\be
\ba
\Delta_{11}^{0}(k)=\Delta_{F}^{0}(k)+\Delta_{T}^{0}(k),
\quad
\quad
\Delta_{22}^{0}(k)=-(\Delta_{11}^{0}(k))^*
\nwl
\Delta_{12}^{0}(k)=-2\pi i\delta(k^2-m^2)m_B(k) ,
\quad
\quad
\Delta_{21}^{0}(k)=\Delta_{12}^{0}(k)
\nwl
\Delta_{F}^{0}(k)=\dsfrac{1}{k^2-m^2+i\varepsilon}
 \quad
\quad
\Delta_{T}^{0}(k)=-2\pi in_B(k)\delta(k^2-m^2),
\nwl
S_{11}^{0}(k)=( \not k+m)(S_{F}^{0}(k)+S_{T}^{0}(k)),
\quad
\quad
S_{22}^{0}(k)=(\not k+m)(S_{F}^{0}(k)+S_{T}^{0}(k))^*
\nwl
S_{12}^{0}(k)=2i\pi (\not k+m)\delta(k^2-m^2)m_F(k)
\quad
\quad
S_{21}^{0}(k)=S_{12}^{0}(k)
\nwl
S_{F}^{0}(k)=\dsfrac{1}{k^2-m^2+i\varepsilon} ,
\quad
\quad
\nwl
S_{T}^{0}(k)=2i\pi\delta(k^2-m^2)\snsq{k_0}.
\lab{amplA25}
\ea
\ee
Here the following notations are used
\be
\ba
\snsq{k_0}=\Theta(k_0)n_F(k)+\Theta(-k_0)\bar {n}_F(k),
\nwl
m_F(k)=\dsfrac{\expon{x/2} \lbr \Theta(k_0)-\Theta(-k_0) \rbr }{\expon{x}+1}=
\dsfrac{sign(k_0)}{2}\sin2\theta_{+k}
\nwl
\quad
m_B(k)=\expon{\beta k_0/2}n_B(k)=\frac{1}{2}\sinh2\theta_k,
\quad
\quad
 x=\beta(k_0-\mu)
\lab{amplA26}
\ea
\ee
where $\Theta$ is the step function :
$\Theta(k_0)=1$ if $k_0 > 0$ and $\Theta(k_0)=0$ otherwise and
$sign(k_0)$ is the sign of $k_0$.

The full propagator satisfies the Dyson equation:
\be
\ba
\Delta_{ab}(k)=\Delta_{ab}^{0}(k)+\ds\sum_{cd}\Delta_{ac}^{0}(k)
\Sigma_{cd}^{B}(k)\Delta_{db}(k)
\nwl
\quad
S_{ab}(k)=S_{ab}^{0}(k)+\ds\sum_{cd}S_{ac}^{0}(k)\Sigma_{cd}^{F}(k)S_{db}(k)
\lab{amplA27}
\ea
\ee
where $\Sigma_{ab}^{F}(k)$ and $\Sigma_{ab}^{B}(k)$ are the proper self
energies of a fermion and boson respectively.
By introducing a complex function $\bar\Sigma_B=Re{\bar\Sigma_B}
  +iIm{\bar\Sigma_B}$ we can represent
the general form of the
self energy  as follows:
\be
\ba
\Sigma_{11}^{B}(k)
=Re{\bar\Sigma_B}(k)+iIm{\bar\Sigma_B}(k)\cosh2\theta_k,
\quad
\quad
\Sigma_{22}^{B}(k)  =-(\Sigma_{11}^{B}(k))^*
\nwl
\quad
\Sigma_{12}^{B}(k)=-i\sinh2\theta_kIm{\bar\Sigma_B}(k),
\quad
\quad
\Sigma_{21}^{B}(k)=\Sigma_{12}^{B}(k) .
\lab{amplA28}
\ea
\ee
Similarly for the  fermion self energy we have:
\be
\ba
\Sigma_{11}^{F}(k) =Re{\bar\Sigma_F}(k)+
i \cos2\theta_k  Im{\bar\Sigma_F}(k)
\quad
\quad
\Sigma_{22}^{F}(k)=(\Sigma_{11}^{F}(k))^*
\nwl
\Sigma_{12}^{F}(k)=-2iIm{\bar\Sigma_F}(k)m_F(k) ,
\quad
\quad
\Sigma_{21}^{F}(k)=\Sigma_{12}^{F}(k).
\lab{amplA29}
\ea
\ee
The full propagator has the  following compact form in terms of
$\bar\Sigma$:
\be
\ba
\Delta_{11}(k)=\dsfrac{ \chsq{k} }{ k^2-m^2-
Re{\bar\Sigma_B}(k)-iIm{\bar\Sigma_B}(k)+i\varepsilon } -
\dsfrac{ \shsq{k} }{ k^2-m^2-
Re{\bar\Sigma_B}(k)+iIm{\bar\Sigma_B}(k)-i\varepsilon }
\nwl
\nwl
\Delta_{12}(k)=m_B(k)\{\dsfrac{ 1 }{ k^2-m^2-
Re{\bar\Sigma_B}(k)-iIm{\bar\Sigma_B}(k)+i\varepsilon } -
\dsfrac{ 1 }{ k^2-m^2-
Re{\bar\Sigma_B}(k)+iIm{\bar\Sigma_B}(k)-i\varepsilon} \}
\nwl
\nwl
\Delta_{22}(k)=-(\Delta_{11}(k))^*,
 \quad
 \quad
 \Delta_{21}(k)=\Delta_{12}(k)
 \lab{amplA30}
 \ea
 \ee
\be
\ba
S_{11}(k)=\dsfrac{ \cssq{k_0} }{ \not p -m-
Re{\bar\Sigma_F}(k)-iIm{\bar\Sigma_F}(k)+i\varepsilon } +
\dsfrac{\snsq{k_0 }     }{ \not p -m-
Re{\bar\Sigma_F}(k)+iIm{\bar\Sigma_F}(k)-i\varepsilon }
\nwl
\nwl
S_{12}(k)=-m_F(k)\{\dsfrac{ 1 }{ \not p -m-
Re{\bar\Sigma_F}(k)-iIm{\bar\Sigma_F}(k)+i\varepsilon } +
\dsfrac{1 }     { \not p -m-
Re{\bar\Sigma_F}(k)+iIm{\bar\Sigma_F}(k)-i\varepsilon } \}
\nwl
\nwl
S_{21}(k)=S_{12}(k)
\quad
\quad
S_{22}(k)=(S_{11}(k))^* \quad ,
\lab{amplA31}
\ea
\ee
where $\chsq{k}=1+n_B$ , $\shsq{k}=n_B(k)$,
 $  \cssq{k_0}=1-\snsq{k_0}  $ and $\snsq{k_0} $ was defined in Eq.
 \re{amplA26}.
The spectral representation for propagators and their matrix form
 may be found in ref.
\ci{saito}.

\bb{99}
\bi{ashok} Ashok Das, Lectures on finite temperature field theory
University of Rochester, 1996;\\
F.C. Khanna and A.E. Santana, Topics in Thermo Field Dynamics,
Lectures given at the second International schooll on field theory
and gravitation. Vitoria,  Brazil, 1999.
\bi{umetak} Y. Takahashi and H. Umezawa, Int. Journ. Mod. Phys.
 10, 1755 (1996), (original paper in Coll. Phenomenon 2, 55 (1975)).
\bi{matsubara} T. Matsubara, Progr. Theor. Phys., {\bf14}, 351 (1955).
\bi{jackiwdolan}L. Dolan and R. Jackiw, Phys. Rev. D {\bf9}, 3320
(1974).
\bi{keldish}J. Schwinger, J. Math. Phys. {\bf2}, 407 (1961);\\
L.V. Keldish, Sov. Phys. JETP, {\bf20}, 1018 (1964).
\bi{niegawa}
 A. Niegawa,
Phys.Rev.D {\bf 57}, 1379 (1998); S. Jeon and P.J. Ellis, Phys. Rev.
{\bf D58}, 04013 (1998); F.T. Brandt, A. Das and J. Frenkel, Phys.
Rev. {\bf D60}, 105008 (1999).
 \bi{feynman} R. Feymnan,  Rev. Mod. Phys. {\bf 20}, 367 (1948).
 \bi{dyson} F. Dyson,  Phys. Rev,  {\bf 75}, 1736 (1949).
 \bi{saito} K. Saito,  T. Maruyama and K. Soutome,  Phys.
 Rev.C {\bf 40}, 407 (1989).
\bi{mandlshaw}  F. Mandl and  G. Shaw,  Quantum field theory.
 (A Wiley - Interscience 1986).
 \bi{cutkosky} F. Cutkosky,  Jour. Math. Phys. {\bf 1}, 429 (1960).
 \bi{kobes} R. Kobes and G. Semenoff,  Nucl. Phys. {\bf B260},  714 (1985);\\
 R. Kobes and G. Semenoff,  Nucl. Phys. {\bf B272},  329 (1986).
 \bi{fujimoto} Y. Fujimoto,  M. Morikawa and M. Sasaki,  Phys. Rev.
 D {\bf 33},
 590 (1986).
 \bi{keil} W. Keil,  Phys. Rev. D {\bf40},  1176 (1989).
 \bi{weldon} H. A. Weldon,  Phys. Rev. D {\bf28}, 2007 (1983).
 \bi{ojima} M. Matsumoto,  I. Ojima and H. Umezawa,  Ann. Phys. (N.Y.)
 {\bf152},  348 (1984).
 \bi{ramond} P. Ramond,   Field theory of modern primer.
 (Benjamin/cumming publishing company) (1981).
 \bi{ourtfd}
A.M. Rakhimov,  U.T. Yakhshev and  F.C. Khanna .
 Phys.Rev.C {\bf 61}, 024907(2000)
\bibitem{songgao} Yi - Jun Zhang,  Song Gao   and Ru - Keng Su
               Phys. Rev. C {\bf 56},  3336 (1997) and references there in.
\bi{zhaokhanna} W. Zhao and F.C. Khanna,  to be published.
\bi{hardy} S.J. Hardy and M.H. Thomas,
 Neutrino - electron processes in a strongly magnetized thermal field,
  astro-ph/0008473,  and references therein.
 \bi{becs}V.G. Bezchastnov and P. Haensel Phys. Rev. D {\bf 54} , 3706 (1996)
 \bi{machelidt}C. Fuchs, I. Senh and H.H. Wolter, Nucl. Phys. A601, 505,
(1996);\\
 R. Machleidt,
 Microscopic calculation of in - medium
 proton - proton cross sections.
 nucl-th/9308016
 \bi{umezawa}
  H.  Umezawa,  H. Matsumoto and H. Tachiki,
  Thermo Field Dynamics and condensed states  (North-
     Holland,   Amsterdam,   1982);  \\
     H. Umezawa,  Advanced field theory,  Micro,  Macro and Thermal
     Physics,  (AIP 1994).
 \eb

\newpage

 \begin{figure}[h]
 \epsfclipon
 \epsfxsize=20.cm
 \centerline{\epsffile{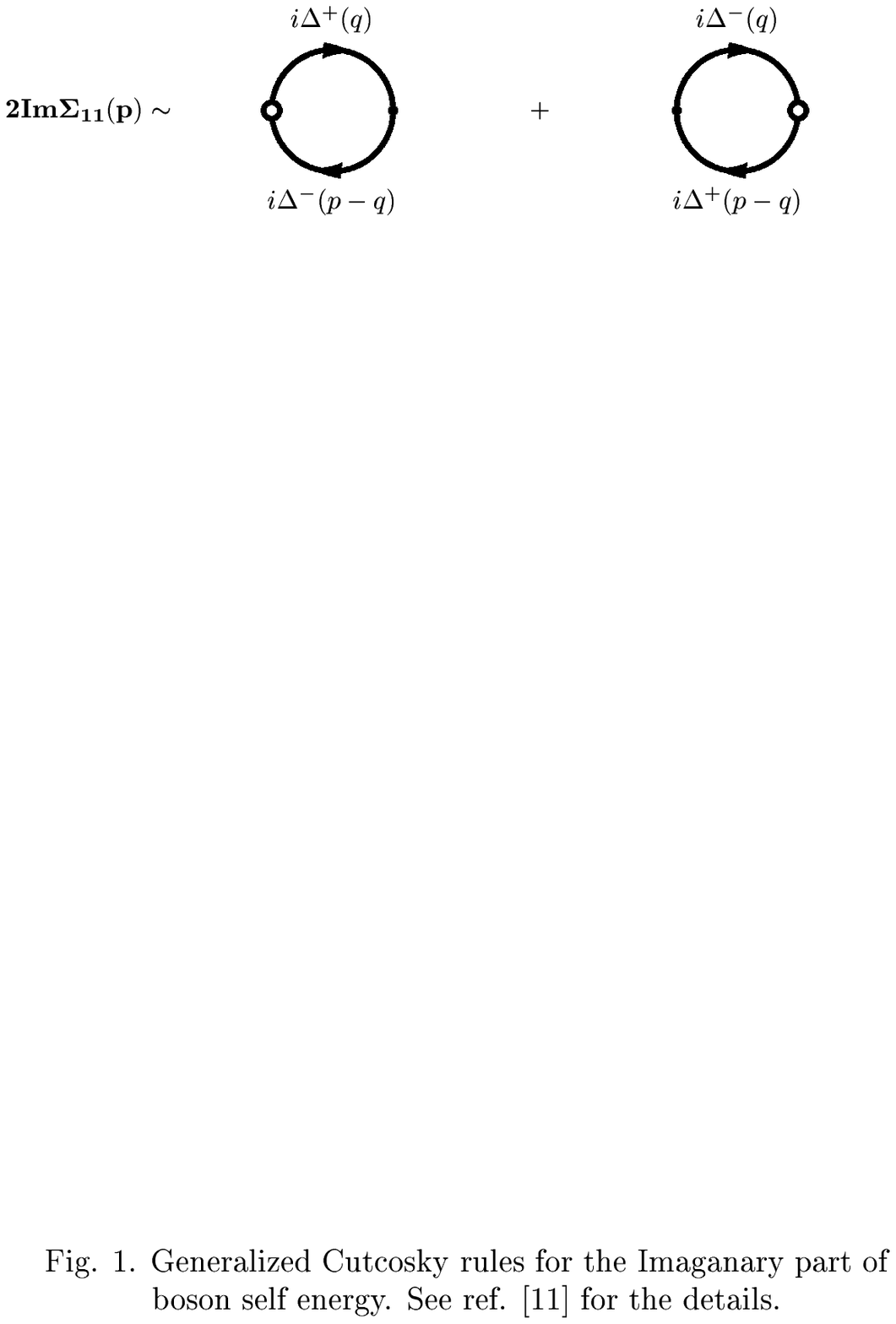}}
 \end{figure}

 \begin{figure}[h]
 \epsfclipon
 \epsfxsize=20.cm
 \centerline{\epsffile{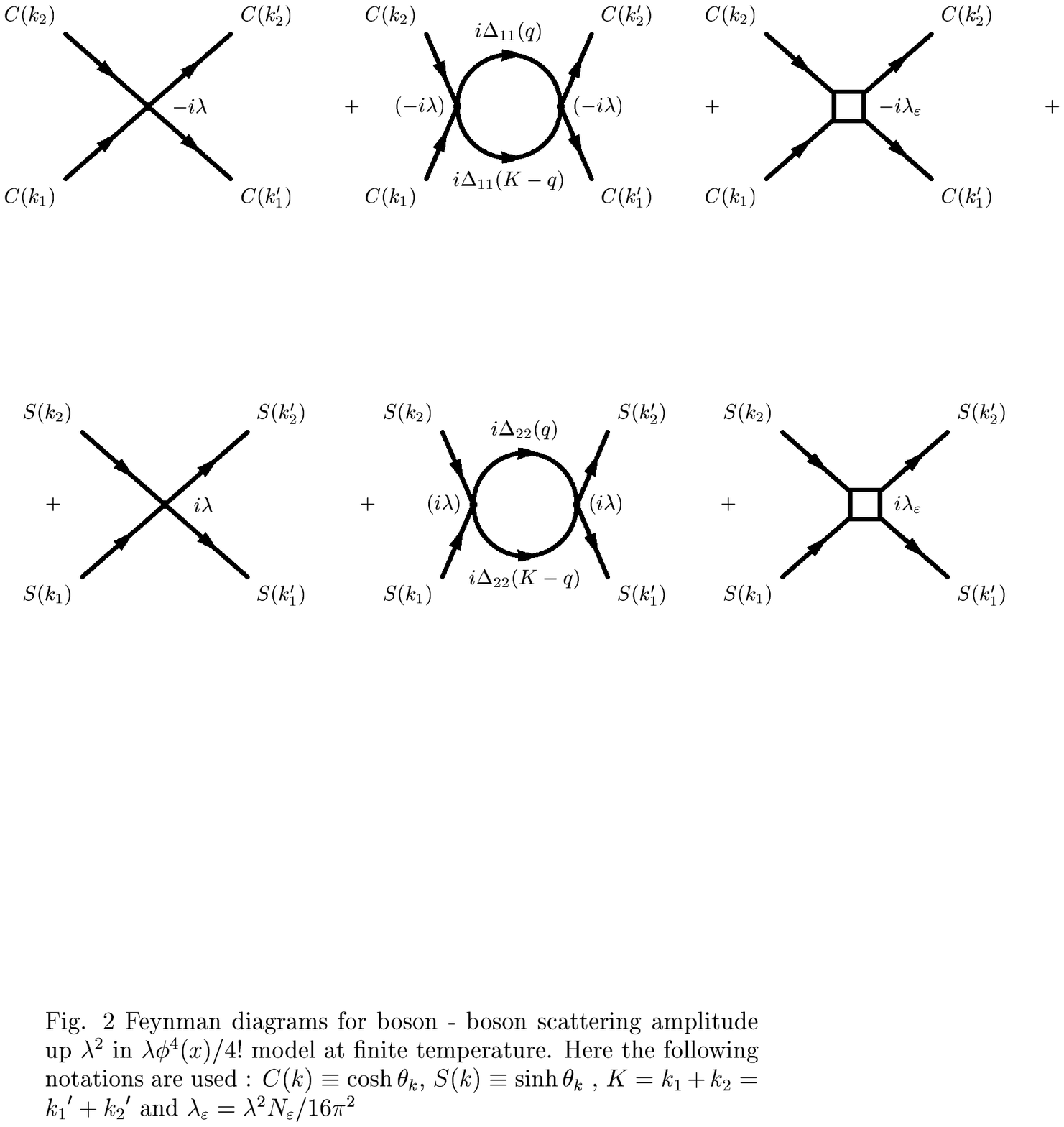}}
 \end{figure}

\end{document}